\documentclass[twocolumn,Prereprint,aps,prb]{revtex4-1}
\usepackage{appendix} 
\usepackage{graphicx}
\usepackage{multirow}
\usepackage{hhline}
\usepackage{comment}
\usepackage{color} 
\usepackage[colorlinks=true, citecolor=green]{hyperref}
\usepackage{float}
\usepackage{soul}
\usepackage{epstopdf}
\usepackage{amsmath}
\usepackage{textcomp}
\usepackage{mathtools}
\usepackage{hyperref}
\usepackage{resizegather}

\DeclarePairedDelimiter\ceil{\lceil}{\rceil}

\begin{document}
\title{Generating Periodic Grain Boundary Structures: Algorithm and Open-Source Python Library}

\author{Jianli Cheng, Jian Luo and Kesong Yang}
\email{kesong@ucsd.edu,+1-858-534-2514}
\affiliation{Department of NanoEngineering, University of California, San Diego, 9500 Gilman Drive, Mail Code 0448, La Jolla, California 92093-0448, USA}

\begin{abstract} 
An algorithm implemented in an open-source python library  was developed for building periodic coincidence site lattice (CSL) grain boundary models  in a universal fashion. The software framework aims to generate tilt and twist grain boundaries from cubic and tetragonal crystals for \textit{ab-initio} and classical atomistic simulation. This framework has two useful features: i) it can calculate all the CSL matrices for generating CSL from a given Sigma ($\Sigma$) value and rotation axis, allowing the users to build the specific CSL and grain boundary models; ii) it provides a convenient command line tool to enable high-throughput generation of tilt and twist grain boundaries by assigning an input crystal structure, $\Sigma$ value, rotation axis, and grain boundary plane. The developed algorithm in the open-source python library is expected to facilitate studies of grain boundary in materials science. The software framework is available on the website: \url{aimsgb.org}.
\end{abstract}

\maketitle
\newpage

\section{INTRODUCTION}
Interfaces are ubiquitous in solid crystalline materials, and can exhibit material properties that are drastically different from their corresponding bulk materials, bringing potential applications in various industrial areas. One example is the discovery of two-dimensional electron gas at perovskite oxide interfaces such as LaAlO$_3$/SrTiO$_3$ heterointerface,\cite{Ohtomo_2004_nature,kesong_2016_sr,Wang_2016_ami,Cheng_2018_jmacc} in which two band insulators combines together and forms highly conducting interfacial conducting states. 
Compared to the heterointerface between two phases with different crystal structures  (or \textit{heterophase interface}),
grain boundaries are one relatively simple but ubiquitous and technologically important interfaces 
that consists of the same phase crystal with different orientations, and thus can be considered as a \textit{homophase interface}.\citep{lejcek2010grain,braccini2013mechanics}

As one common planar defect, grain boundaries can control microstructural evolution and significantly change the mechanical, electronic, and other properties of polycrystalline materials.
\citep{Mishin_2010_actamat,Karma_2012_prl,van_2013_nmat, Behtash_2018_jacers, Huang_2014_actamat} 
Investigating the influence of grain boundaries on the materials properties has been an important research subject in materials science.\citep{Gellert_2012_jpcc,Breuer_2015_jmca,Bowman_2017_nanoscale}
For instance, grain boundaries in solid electrolytes can often have ionic conductivities that are several orders of magnitude lower than that of bulk, thereby limiting the overall ionic conductivity and severely lowering the device performance.\cite{Gellert_2012_jpcc,Breuer_2015_jmca,Bowman_2017_nanoscale,Ma_2014_ees}
In addition, grain boundaries can undergo structural transformations, which can abruptly change their mobility (grain growth rate and microstructural development) and mechanical properties of polycrystalline materials,
though direct experimental evidences for these phenomena are lacking due to extreme challenges of grain boundary structure characterization at high temperature from high-resolution transmission electron microscopy.\cite{Frolov_2013_ncom,Frolov_2013_prl,Karma_2012_prl}
Therefore, to characterize a grain boundary structure, particularly from a computational modeling viewpoint, is of great importance to understand the influence of grain boundaries on the materials properties, and to design high-performance engineering materials.

To describe a grain boundary from crystallography, one needs five macroscopic degrees of freedom (DOFs).\citep{Sutton_1995} 
The five DOFs define how the two grains with required orientation are combined to form a grain boundary from a given crystal structure.
Three of them describe  mutual misorientations between two adjoining grains, which are represented by a  rotation axis \textbf{\textit{o}} (two DOFs) and a rotation angle $\theta$ (one DOF). The remaining two DOFs describe the normal \textbf{\textit{n}} to the grain boundary plane, which indicates the orientation of the grain boundary between the two disorientated grains. The geometrical alignment between  \textbf{\textit{o}} and \textbf{\textit{n}} defines the grain boundary type:
tilt (\textbf{\textit{o}}$\bot$ \textbf{\textit{n}}), twist (\textbf{\textit{o}} $\parallel$ \textbf{\textit{n}}), and mixed grain boundaries. \citep{Sutton_1995}

Among all the types of grain boundaries, there exist one type of special grain boundaries called \textit{coincidence-site lattice} (CSL) grain boundary. In the CSL grain boundary, some atomic sites of one grain coincide exactly with some atomic sites of the other grain, and these special atomic sites are called \textit{coincidence sites}. The \textit{coincidence sites} are spread periodically throughout the whole superimposition and create a supercell called CSL.\citep{Bollmann_1970, lejcek2010grain} 
Compared to random (non-special) grain boundaries, the CSL grain boundaries are believed to have low grain boundary energy because of good atomic fit,\citep{Brandon_1964_ActaM, Brandon_1966_ActaM,Pan_1994_SMM_CSL, Sangid_2010_MSEA_GB} 
and they have been one important research topic in the grain boundary science and engineering.
Given the enormous complexity of grain boundaries, an efficient algorithm to build any CSL grain boundary structures from a minimum user input is of great usage for the structural characterization of grain boundaries. 
In particular, as the  emergence of high-throughput computational techniques and materials informatics,\cite{kesong_2012_NM,Curtarolo_2013_NMAT,Rajan_2005_MI_MToday} such a tool is very necessary to facilitate high-throughput computational studies of the grain boundaries. 

In spite of some softwares available to construct grain boundary,\cite{Ogawa_2006_mt, Palmer_2015_crystalmaker}
to the best of our knowledge, 
 there is no universal and easy-to-use tool to generate grain boundary structures in a high-throughput fashion. 
For instance, Ogawa developed a web-based applet called as \textit{GBstudio} to generate periodic grain boundary structures with $\Sigma$ value up to 99.\cite{Ogawa_2006_mt} 
The \textit{GBstudio} requires users to build primitive crystal structure online first and then generate grain boundary structures for a given $\Sigma$ value. This process of generating grain boundaries is extremely laborious and even impossible for a complex crystal structure such as organic-inorganic hybrid halide perovskite CH$_3$NH$_3$SnI$_3$.\citep{Bernal_2014_jpcc} 
It also makes it extremely difficult to work with high-throughput calculations.
Another graphical user interface (GUI) software to build grain boundary is \textit{CrystalMaker},\cite{Palmer_2015_crystalmaker} 
which requires users to build slab or surfaces first and then combine them as a grain boundary structure.
As in the case of \textit{GBstudio}, this GUI-based nature makes it incapable of high-throughput production of grain boundaries. Unlike the GUI-based softwares, some species of code based on the command line interface were also developed to generate grain boundaries, which are either limited to several crystal structures only or with limited functions.\citep{Wojdyr_2010_msmse, Plimpton_1995} 
Also the users are often required to modify the source code or to add new functions to build desired grain boundaries.   

In this article, we introduce one efficient algorithm to build periodic grain boundary models for \textit{ab-initio} and classical atomistic simulation in materials science. The algorithm is implemented in an open-source python library that aims to generate tilt and twist grain boundaries from a cubic or tetragonal crystal in a universal fashion.  The code is freely available for download at \url{aimsgb.org}.

\section{Building Procedures}
Here is a brief outline of the algorithm to build grain boundary.

(1) Calculate rotation angle for a given $\Sigma$ value and rotation axis.

(2) Generate a rotation matrix from the rotation angle and rotation axis.

(3) Calculate CSL matrix using the rotation matrix and $\Sigma$ value. 

(4) Generate two grains from CSL matrix, and combine them based on a given grain boundary plane to build grain boundary.

\subsection{Rotation Angle}
Grain boundary represents one special interface that consists of two grains of the same phase, unlike the case of heterointerface between two phase structures.\cite{kesong_2016_sr,Wang_2016_ami,Cheng_2018_jmacc} 
The two grains differ in their mutual orientations and this misorientation can be described by  a rotation that brings the two grains into coincidence. 
The rotation is characterized by a rotation axis \textbf{\textit{o}} [$uvw$] and a rotation angle $\theta$. 
In the CSL theory, the geometry of a grain boundary structure is described by an integer number ($\Sigma$), which is traditionally defined as a ratio between the volume enclosed by a CSL (\textit{coincidence unit cell volume}) and 
unit cell volume of a cubic crystal with a rotation along [001] rotation axis.\citep{Priesteri_2012} Here, by including a rotation with any rotation axis, we redefine the $\Sigma$ as a ratio between the \textit{coincidence unit cell volume} and that of a rotated unit cell of crystal (\textit{rotated unit cell volume}):
\begin{equation}
\Sigma = \frac{Coincidence{\;}unit{\;}cell{\;}volume}{Rotated{\;}unit{\;}cell{\;}volume}
\label{eqsigma1}
\end{equation}
The \textit{rotated unit cell volume} depends on the rotation axis \textit{\textbf{o}} [$uvw$], and equals to ($u^2 + v^2 + w^2$) times of the volume of the conventional unit cell for a cubic lattice. 
In other words, the eq. \ref{eqsigma1} can be rewritten as:
\begin{equation}
\Sigma = \frac{Coincidence{\;}unit{\;}cell{\;}volume}{Conventional{\;}unit{\;}cell{\;}volume \times (u^2 + v^2 + w^2)}
\label{eqsigma2}
\end{equation}
For example, a rotated unit cell with a [001] rotation axis has the same volume with its conventional unit cell, while a rotated unit cell with a [111] rotation axis has a volume of three times of the conventional unit cell.
Another equivalent definition of  $\Sigma$ is the ratio of the total number of sites in the coincidence unit cell to that of the rotated unit cell. Note that in some special grain boundaries, like $\Sigma$5[001]/(130), the ratio is an even number 10, and in this case, the $\Sigma$ will be half of the ratio. Also, if the $\Sigma$ is a multiple of $u^2 + v^2 + w^2$, the generated grain boundary will be same with a $\Sigma/(u^2 + v^2 + w^2)$ grain boundary. For example, $\Sigma$21[111] is same with $\Sigma$7[111].

\begin{figure}[t]
\centering
\includegraphics[width=0.49\textwidth]{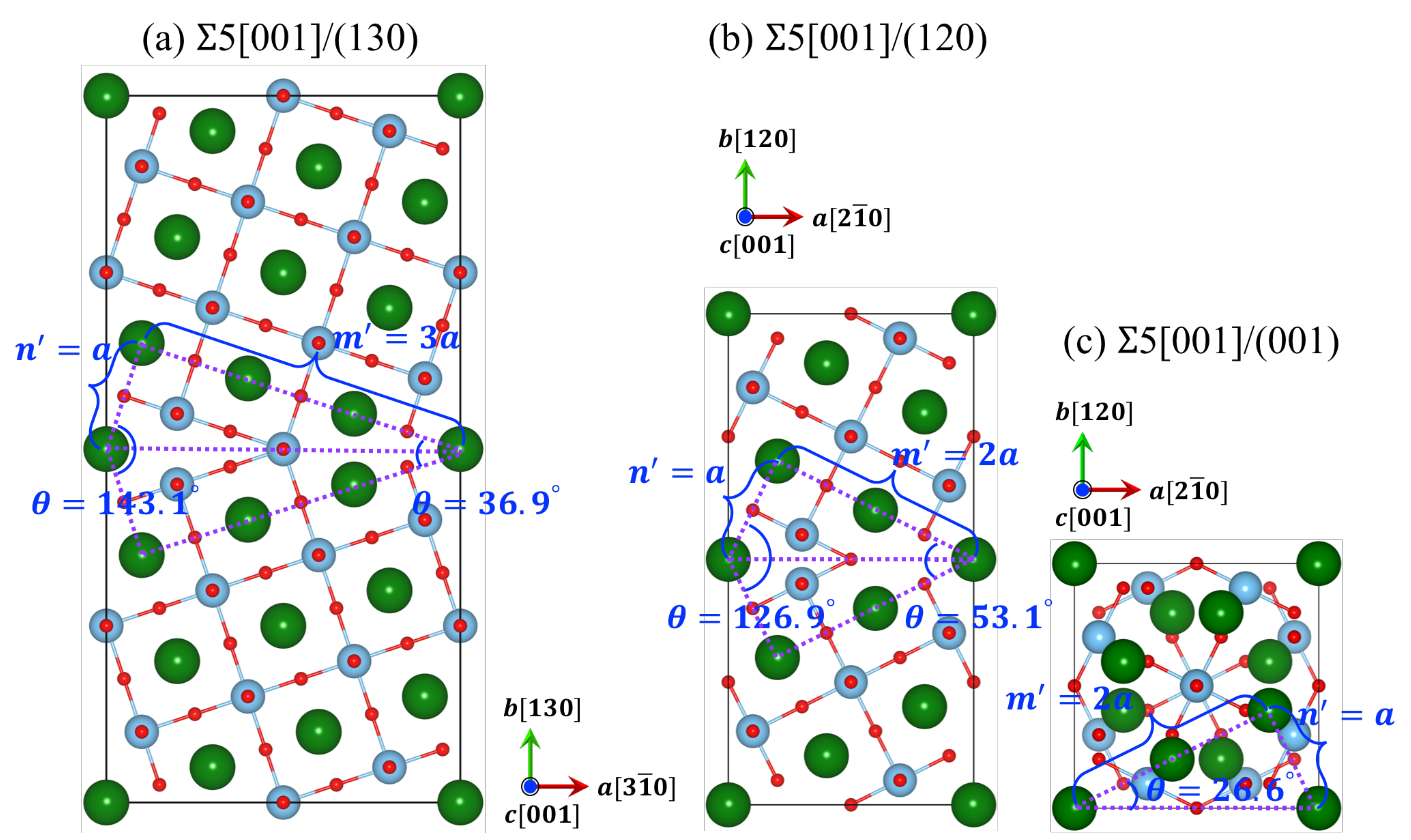}
\caption{ $\Sigma$5[001] grain boundary structures of cubic SrTiO$_3$. (a) $\Sigma$5[001]/(130), (b) $\Sigma$5[001]/(120), and (c) $\Sigma$5[001]/(001). 
The former two structures are tilted grain boundaries, and the third one is twisted grain boundary in this and subsequent two figures.
}
\label{sto_001_5}
\end{figure}

To generate a CSL from a conventional unit cell of a crystal, $\Sigma$, rotation axis \textbf{\textit{o}} [$uvw$], and rotation angle $\theta$ must satisfy the following conditions:\citep{Grimmer_1984_actaca, Bleris_1981_actaca, Ranganathan_1966_actacryst}
\begin{equation}
A = \alpha\Sigma = m^2 + (u^2 + v^2 + w^2)n^2
\label{eqsigma}
\end{equation}
and
\begin{equation}
tan(\frac{\theta}{2}) = \frac{n}{m}(u^2 + v^2 + w^2)^{1/2}
\label{eqtheta}
\end{equation}
where $\alpha$ = 1, 2 or 4,\citep{Bleris_1981_actaca} $m$ and $n$ are positive integers. Thus, for a given $\Sigma$, the range of $m$ and $n$ can be determined as 1 $\leq m \leq \ceil*{2\sqrt{\Sigma}}$ and 0 $\leq n \leq \ceil*{2\sqrt{\Sigma}}$, respectively, with $\ceil*{2\sqrt{\Sigma}}$ mapping the least integer numbers greater than or equal to 2$\sqrt{\Sigma}$. 
In addition, the integers $u$, $v$, $w$ and the integers $m$, $n$ should be coprime, \textit{i.e.}, with greatest common divisor (GCD) equal to 1,
\begin{equation}
\text{GCD}(u, v, w) = 1 \qquad\text{and}\qquad \text{GCD}(m, n) = 1
\label{eq3}
\end{equation}

Accordingly, by assigning \textbf{\textit{o}} and $\Sigma$, one can determine the values of $m$ and $n$ from eq. \ref{eqsigma} and \ref{eq3}. 
It is noted that for each $\Sigma$, there could be multiple sets of ($m$, $n$) values and each set corresponds to a different rotation angle $\theta$. 
Here we take SrTiO$_3$ $\Sigma5$ [001] grain boundary (see Fig. \ref{sto_001_5}) as an example to show the process of deducing rotation angle and corresponding (\textit{m}, \textit{n}) values.
For $\Sigma5$ [001] grain boundary, eq. \ref{eqsigma} gives $5\alpha = m^2 + n^2$, with $\alpha$ = 1, 2, or 4. 
Note that herein $m^2 + n^2$ also represents the ratio between the area enclosed by a unit cell of CSL and the unit cell of the standard conventional crystal  structure. By using eqs. \ref{eqsigma} and \ref{eqtheta}, we are able to conclude that: 

(1) At $\alpha$ = 1, ($m$, $n$) = (2, 1) and $\theta$ = 53.1$^{\circ}$, or ($m$, $n$) = (1, 2) and $\theta$ = 126.9$^{\circ}$. The area of generated CSL is five times as large as the standard conventional unit cell. Note that the two  rotation angles, $\theta$ = 53.1$^{\circ}$ and 126.9$^{\circ}$, correspond to the same grain boundary since their sum equals to 180$^{\circ}$.

(2) At $\alpha$ = 2, ($m$, $n$) = (3, 1) and $\theta$ = 36.9$^{\circ}$, or ($m$, $n$) = (1, 3) and $\theta$ = 143.1$^{\circ}$. The area of each CSL is 10 times as large as the standard conventional unit cell. The two rotation angles here also  correspond to the same grain boundary.

(3) At $\alpha$ = 4, ($m$, $n$) = (4, 2) or (2, 4), the CSL generated in this case is the same with that in $\alpha$ = 1, as shown from eq. \ref{eq3}.

\begin{figure}[t]
\bigskip
\includegraphics[width=0.45\textwidth]{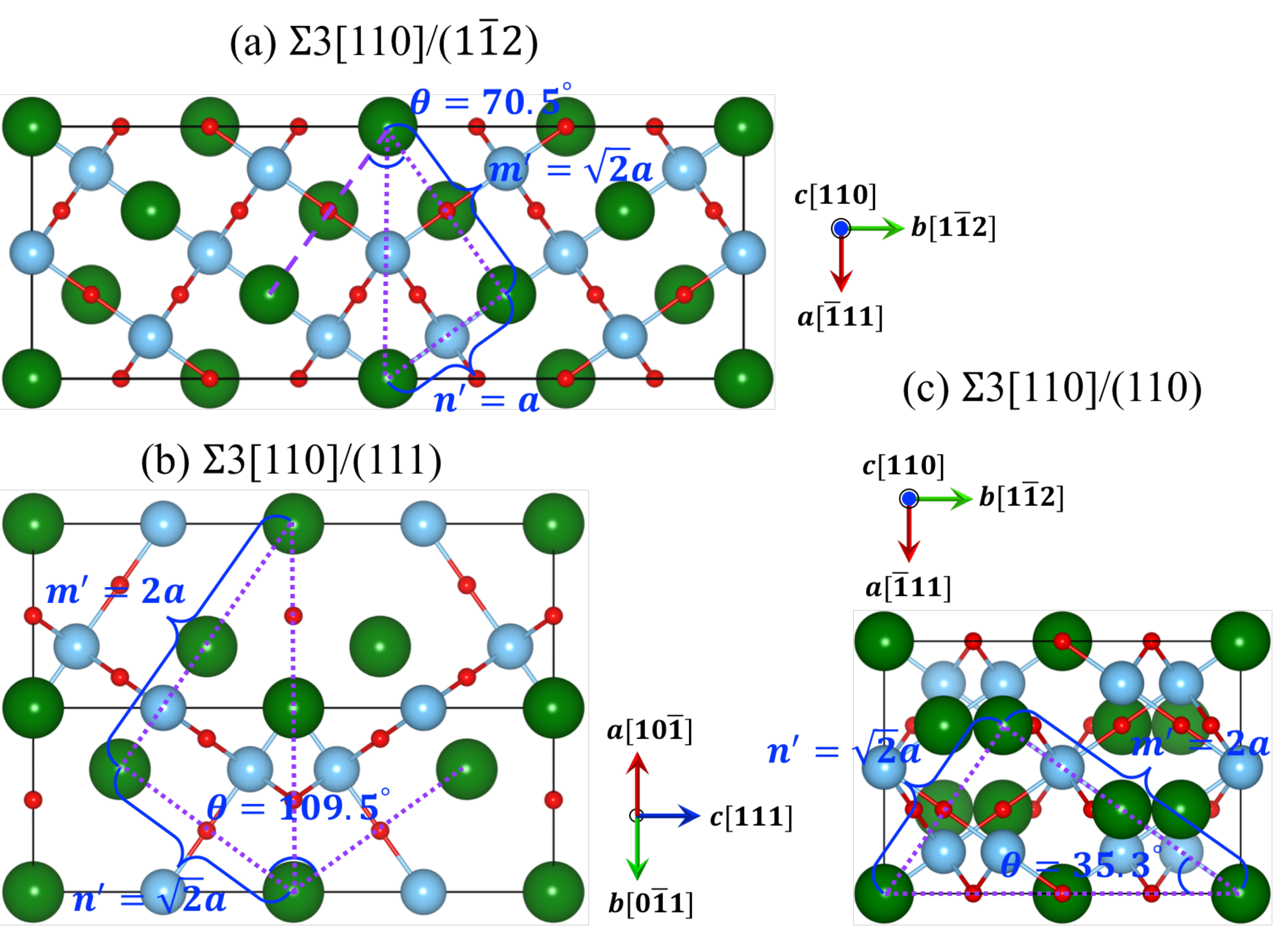}
\caption{$\Sigma$3[110] grain boundary structures of cubic SrTiO$_3$. (a) $\Sigma$3[110]/(1$\bar{1}$2), (b) $\Sigma$3[110]/(111), and (c) $\Sigma$3[110]/(110). 
$\Sigma$3[110]/(111) is reduced primitive structure in which the angle between axis-$a$ and $b$ equals to 120 $^{\circ}$.
}
\label{sto_110_3}
\end{figure}

\begin{figure}[t]
\includegraphics[width=0.45\textwidth]{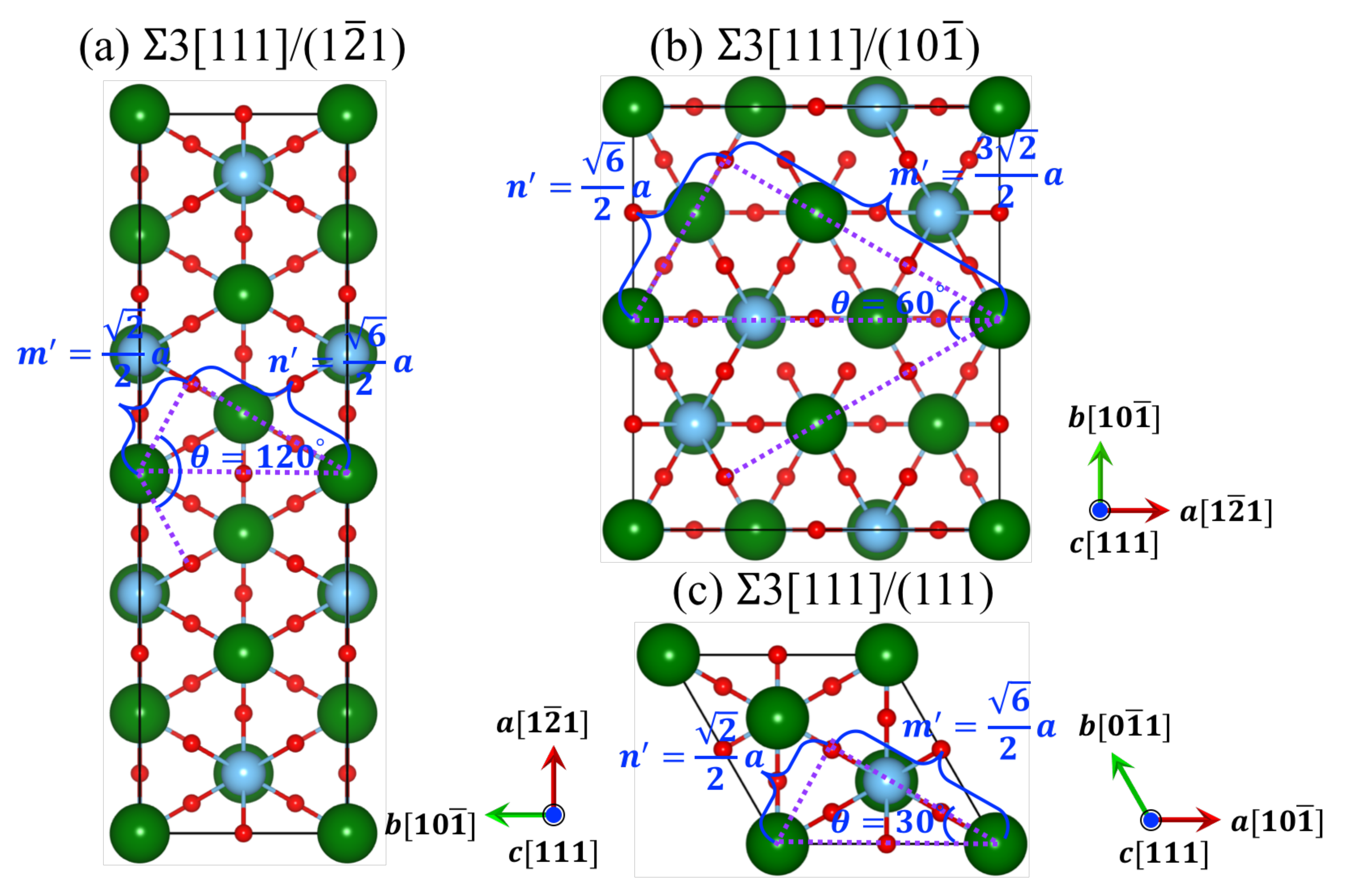}
\caption{$\Sigma$3[111] grain boundary structures of cubic SrTiO$_3$. (a) $\Sigma$3[111]/(1$\bar{2}$1), (b) $\Sigma$3[111]/(10$\bar{1}$), and (c) $\Sigma$3[111]/(111). 
$\Sigma$3[111]/(111) is reduced primitive structure in which the angle between axis-$a$ and $b$ equals to 120 $^{\circ}$.}
\label{sto_111_3}
\end{figure}

As discussed later, in addition to the rotation axis \textbf{\textit{o}} and rotation angle $\theta$, another degree of freedom is the grain boundary plane. 
For example,  $\Sigma5$ [001] grain boundary has three grain boundary planes, (130), (120), and (001), see Fig. \ref{sto_001_5}. The grain planes (130) and (120) are parallel to the rotation axis [001] and belong to tilted grain boundaries, while the grain plane (001) is perpendicular to the rotation axis [001] and belongs to twisted grain boundary. In addition, to have a clear presentation of the building procedure, we also show the $\Sigma3$ [110] and $\Sigma3$ [111] grain boundary structures of cubic SrTiO$_3$ in Fig. \ref{sto_110_3} and Fig. \ref{sto_111_3}, respectively.

\subsection{Rotation Matrix}
After getting $\theta$ from a given $\Sigma$ and \textbf{\textit{o}}, we are able to generate a rotation matrix (\textbf{R}) using the Rodrigues' rotation formula.\citep{Priesteri_2012}
First, we define a unit vector \textbf{\textit{k}} = $[k_x, k_y, k_z]$, where  ${k_x}^2$+${k_y}^2$+${k_z}^2$=1, which represents the rotation axis \textbf{\textit{o}} [uvw], and a matrix \textbf{\textit{K}} that denotes the cross-product matrix for the unit vector \textbf{\textit{k}}.

\begin{gather}
 \textbf{\textit{K}}
 =
  \begin{bmatrix}
   0 & -k_z & k_y \\
   k_z & 0 & -k_x \\
   -k_y & k_x & 0
   \end{bmatrix}
\label{eq5}
\end{gather}
Accordingly, a matrix \textbf{\textit{R}}  that describes a rotation with an angle  $\theta$ in counterclockwise around the axis \textbf{\textit{k}} can be given as:
\begin{equation}
\textbf{\textit{R}} = \textbf{I} + (sin\theta)\textbf{\textit{K}} + (1 - cos\theta)\textbf{\textit{K}}^2
\label{eq6}
\end{equation}
where \textbf{I} is a 3$\times$3 identity matrix.

\subsection{CSL Matrix}

\begin{figure}[t]
\centering
\includegraphics[width=0.45\textwidth,clip]{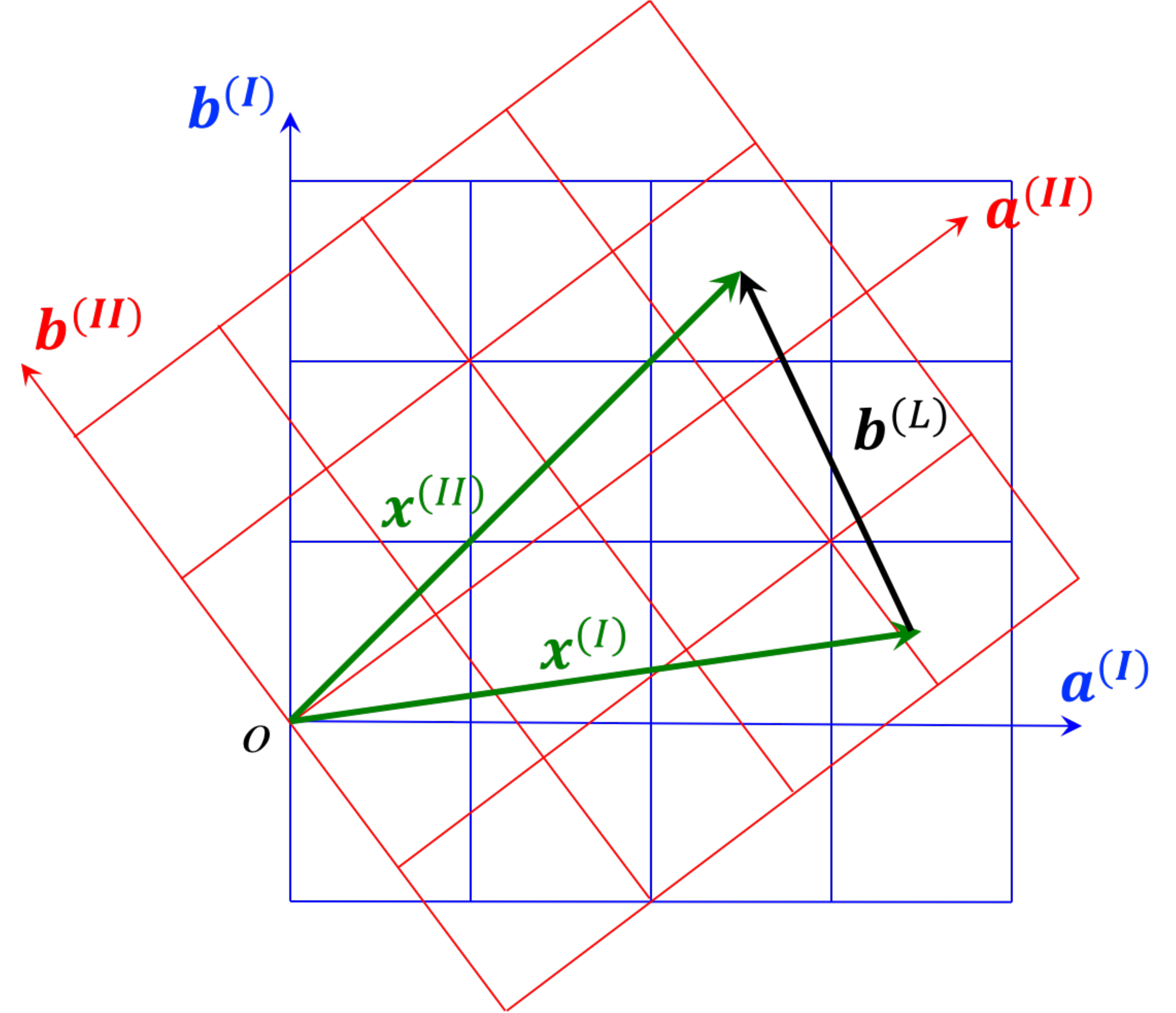}
\caption{Schematic illustration of general operation on coincidence point in a two-dimensional space.}
\label{o_lat_theory}
\end{figure}

Next step is to calculate CSL matrix from the rotation matrix \textbf{R}. 
Here, we implement a so-called \textit{O-lattice} theory introduced by Bollmann,\citep{Bollmann_1970} 
which is an effective tool to analyze the coincidence lattice points of two interpenetrating misoriented lattices of grains. 
The schematic illustration of applying \textit{O-lattice} in deviating equivalence points in two lattices in a two-dimensional space is shown in Fig. \ref{o_lat_theory}.
Assume that we have lattice \textbf{I} with basis lattice vectors \textbf{\textit{a}$^{(I)}$} and \textbf{\textit{b}$^{(I)}$}, and lattice \textbf{II} with basis lattice vectors \textbf{\textit{a}$^{(II)}$} and \textbf{\textit{b}$^{(II)}$}. 
An arbitrary point within the elementary cell of lattice \textbf{I} can be described by a vector \textbf{\textit{x}$^{(I)}$}.
\textbf{\textit{x}$^{(I)}$} can be transformed into a vector \textbf{\textit{x}$^{(II)}$} in lattice \textbf{II} by applying a transformation matrix \textbf{A}:
\begin{equation}
\textbf{\textit{x}$^{(II)}$} = \textbf{A\textit{x}$^{(I)}$}
\label{eq7}
\end{equation}
Therefore, the lattice point (\textbf{\textit{x}$^{(II)}$}) within the lattice \textbf{II} is an equivalent point to the one (\textbf{\textit{x}$^{(I)}$}) in crystal lattice \textbf{I}. 
Moreover, the coincidence point \textbf{\textit{x}$^{(II)}$} also belong to both lattice \textbf{I}, and thus it can also be obtained from lattice \textbf{I} by adding a translation vector \textbf{\textit{t}$^{(L)}$} to \textbf{\textit{x}$^{(I)}$}:
\begin{equation}
\textbf{\textit{x}$^{(II)}$} = \textbf{\textit{x}$^{(I)}$} + \textbf{\textit{t}$^{(L)}$}
\label{eq8}
\end{equation}
By combining eq. \ref{eq7} and \ref{eq8}, one can get:
\begin{equation}
(\textbf{I} - \textbf{A}^{-1})\textbf{\textit{x}$^{(II)}$} = \textbf{\textit{t}$^{(L)}$}
\label{eq9}
\end{equation}
where \textbf{I} is the identity matrix, \textit{i.e.}, unit transformation.
Let us name all the coincidence points as \textbf{\textit{x}$^{(0)}$}, then the eq. \ref{eq9} can be rewritten as:
\begin{equation}
(\textbf{I} - \textbf{A}^{-1})\textbf{\textit{x}$^{(0)}$}  = \textbf{T\textit{x}$^{(0)}$}  = \textbf{\textit{t}$^{(L)}$}
\label{eq10}
\end{equation}
\begin{equation}
\textbf{I} - \textbf{A}^{-1} =  \textbf{T} 
\label{eq10B}
\end{equation}

The solution to this equation requires that $\begin{vmatrix}\textbf{I} - \textbf{A}^{-1}\end{vmatrix}\neq0$.
For convenience, we hereafter apply transformations  in the conventional crystal coordinate system.
The corresponding transformation matrices in the crystal coordinate system are labeled as  \textbf{A$^{\prime}$}  and \textbf{T$^{\prime}$}, and the following expressions are to be calculated:\citep{Grimmer_1974_actaca}
\begin{equation}
\textbf{I} - \textbf{A}^{\prime -1} = \textbf{T}^{\prime} = \textbf{I} - \textbf{US$^{-1}$R$^{-1}$S}
\label{eq13}
\end{equation}
\begin{equation}
\text{det}(\textbf{T}^{\prime}) = \frac{n}{\Sigma}
\label{eq14}
\end{equation}
\begin{equation}
{\textbf{X}^{\prime}}^{(0)} = {\textbf{T}^{\prime}}^{-1}
\label{eq15}
\end{equation}
where \textbf{R}  is rotation matrix as calculated before; \textit{n} is an integer number calculated from eq. \ref{eq14}; ${\textbf{X}^{\prime}}^{(0)}$ is a matrix whose three column vectors are unit vectors of the O-lattice; \textbf{S} is the structure matrix which contains the unit vectors of the crystal coordinate system expressed in the orthonormal coordinates of the conventional crystal, and \textbf{S} = \textbf{I}  for conventional cubic structure; \textbf{U} is a unimodular transformation matrix (det(\textbf{U})=$\pm$1) that is used to keep det($\textbf{T}^{\prime}$) $\neq$ 0.

After obtaining ${\textbf{X}^{\prime}}^{(0)}$, next we are able to determine CSL matrix  $\textbf{C}^{\prime}$ (in crystal coordinate system). Note that  $\textbf{C}^{\prime}$ should always consist of integer numbers and can be determined as two steps: i) operate on two columns of ${\textbf{X}^{\prime}}^{(0)}$ to let them become integers and meanwhile keep their determinant unchanged; ii) multiply the third column by \textit{n} so that the determinant of matrix $\textbf{C}^{\prime}$ equals to $\Sigma$.
The calculated CSL matrix $\textbf{C}^{\prime}$ can be further reshaped so that each of its column vectors has the shorted length. Moreover, the CSL matrix $\textbf{C}^{\prime}$ will be orthogonalized, with the third column vector same as the rotation axis. Note that the generated CSL matrix will be applied on a cubic lattice and therefore the CSL matrix has one important property. That is, the three numbers in each column of the CSL matrix correspond to a group of miller indices that represent a surface plane as well as surface normal  since basis lattice vectors of the input crystal are orthogonal, as shown in the appendix table of CSL matrix. 
Also note that the CSL matrix can also be applied on a tetragonal lattice if the rotation axis is normal to the tetragonal plane, such as tetragonal TiO$_2$ lattice with [001] rotation.

\subsection{Create Grain Boundary}

A grain boundary consists of two grains that are symmetrical with the grain boundary plane as the mirror plane.\cite{lejcek2010grain} The orientation of a grain boundary plane is given by a unit vector \textit{\textbf{n}}, \textit{i.e.}, the normal to the grain boundary plane.\cite{Pavel_CRSSMS_1995_GB}
Using the generated CSL matrix, we can build a new crystal lattice that is one grain of the grain boundary (named as grain A). The other grain of the grain boundary (named as grain B) can be obtained by applying mirror symmetry operation on grain A. 
After getting two grains A and B, we can build a grain boundary by combining the two grains. As discussed above, the three numbers in each column vector of a CSL matrix indicate a surface plane (also a surface normal, \textit{i.e.}, a direction), and the third column of the CSL matrix is set same with the rotation axis. As a result, the generated new crystal lattice from the CSL matrix has three surface planes (corresponding to three columns of a CSL matrix) and each of them can serve as a grain boundary plane. Note that the relationship between the rotation axis (\textit{\textbf{o}}) and the normal to the grain boundary plane (\textbf{\textit{n}}) determines the type of generated grain boundaries: \textit{tilt} grain boundary for \textbf{\textit{o}} $\bot$ \textbf{\textit{n}} and \textit{twist} grain boundary for \textbf{\textit{o}} $\parallel$ \textbf{\textit{n}}.
Therefore, we are able to generate two \textit{tilt} grain boundaries  by setting first two surface planes as the grain boundary plane, and one \textit{twist} grain boundary by setting third surface plane as the grain boundary plane. 
In other words, one CSL matrix corresponds to three grain boundaries.

\begin{figure}
\centering
\includegraphics[width=0.47\textwidth,clip]{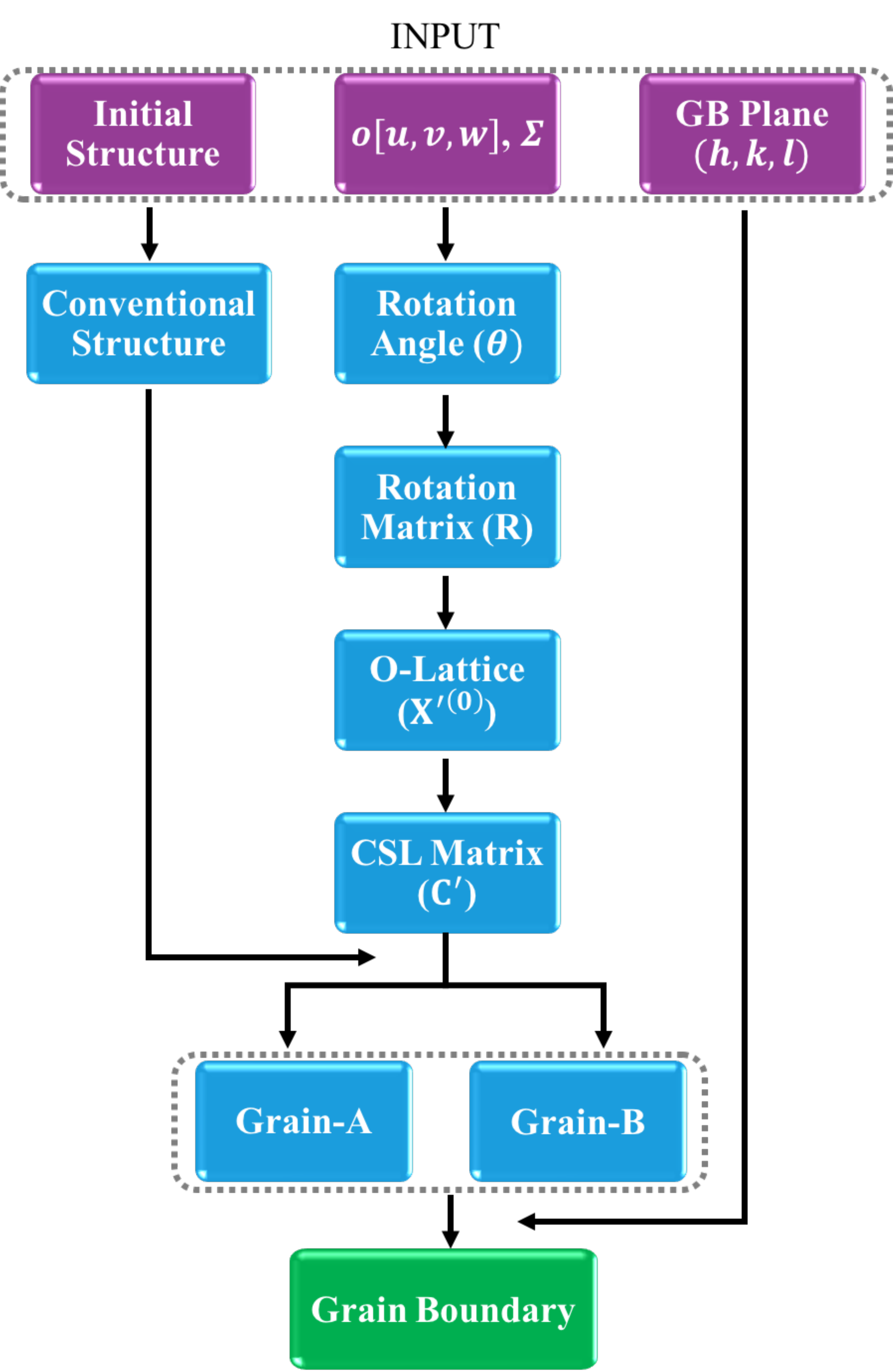}
\caption{The complete building procedure for generating atomic coordinates of periodic grain boundary models from the input $\Sigma$, rotation axis, grain boundary plane and initial crystal structure}
\label{build_process}
\end{figure}

By summarizing the above building procedures, we show the complete workflow in Fig. \ref{build_process}. 
The algorithm is implemented in an open-source python library, aimsgb. 
The representation and manipulation of structures are treated through either pymatgen\cite{Ping_2013_cms} or structural class of AFLOW.\cite{aflowPAPER}
Also, to demonstrate the efficiency of our algorithm, we list the required CPU time (without counting time to write to files) to generate grain boundary structures  for SrTiO$_3$, cubic CH$_3$NH$_3$SnI$_3$, and tetragonal anatase TiO$_2$,  see Table \ref{table1}. The total number of atoms in the grain boundary structure was also listed for each $\Sigma$. As a comparison, the total time  required to generate grain boundary structures, including the time to write to files,  are listed in appendix table.

\begin{figure*}
\centering
\includegraphics[width=0.8\textwidth,clip]{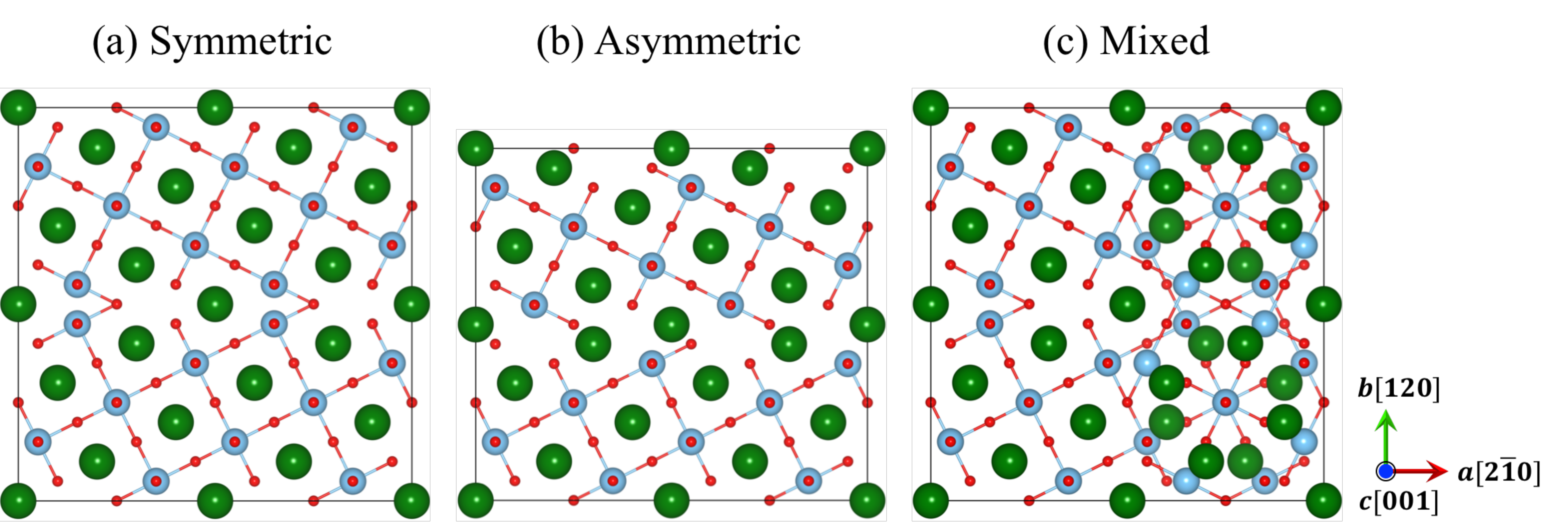}
\caption{Schematic illustrations of (a) symmetric, (b) asymmetric, and (c) mixed grain boundary models.}
\label{asy_mix}
\end{figure*}

Moreover, in addition to the symmetric tilt and twist grain boundaries, aimsgb can also generate asymmetric tilt, twist, and even mixed grain boundaries by modifying the interfacial terminations of the two grains, as these asymmetric  models may represent more realistic polycrystalline systems.\citep{Mishin_2010_actamat, Amouyal_2005_actamat, Medlin_2001_actamat, Janssens_2006_nmat} As a proof of the concept, we show structural illustrations of asymmetric and mixed SrTiO$_3$ grain boundaries from the symmetric  $\Sigma$5[001]/(120) grain boundary model in Fig. \ref{asy_mix}.

\begin{table}[t]
\small\addtolength{\tabcolsep}{4.2pt}
\caption{Required CPU time (in second) to generate grain boundary structures for SrTiO$_3$, cubic CH$_3$NH$_3$SnI$_3$, and tetragonal anatase TiO$_2$, without counting the time to write to files. The total number (No.) of atoms in the grain boundary structure is listed for each $\Sigma$.}
\begin{tabular}{c|c|c|c|c|c|c}
\hline
\hline
& \multicolumn{2}{c|}{SrTiO$_3$}& \multicolumn{2}{c|}{CH$_3$NH$_3$SnI$_3$}& \multicolumn{2}{c}{TiO$_2$}\\
\hline
\multirow{2}{*}{$\Sigma$} & No. of & Time & No. of & Time & No. of & Time \\
& atoms & (sec) & atoms & (sec) & atoms & (sec)\\
\hhline{-------}
5	&50	&0.05	&120	&0.06	&120	&0.06	\\
13	&130	&0.02	&312	&0.04	&312	&0.04	\\
17	&170	&0.02	&408	&0.05	&408	&0.05	\\
25	&250	&0.03	&600	&0.07	&600	&0.07	\\
29	&290	&0.04	&696	&0.08	&696	&0.08	\\
37	&370	&0.04	&888	&0.10	&888	&0.10	\\
41	&410	&0.06	&984	&0.11	&984	&0.11	\\
53	&530	&0.06	&1272	&0.14	&1272	&0.14	\\
61	&610	&0.07	&1464	&0.16	&1464	&0.17	\\
65	&650	&0.07	&1560	&0.16	&1560	&0.16	\\
73	&730	&0.09	&1752	&0.20	&1752	&0.20	\\
85	&850	&0.09	&2040	&0.23	&2040	&0.22	\\
89	&890	&0.10	&2136	&0.23	&2136	&0.23	\\
97	&970	&0.12	&2328	&0.26	&2328	&0.27	\\
101	&1010	&0.10	&2424	&0.24	&2424	&0.25	\\
109	&1090	&0.14	&2616	&0.30	&2616	&0.31	\\
113	&1130	&0.12	&2712	&0.28	&2712	&0.28	\\
125	&1250	&0.14	&3000	&0.34	&3000	&0.35	\\
137	&1370	&0.18	&3288	&0.38	&3288	&0.37	\\
145	&1450	&0.16	&3480	&0.36	&3480	&0.36	\\
149	&1490	&0.17	&3576	&0.39	&3576	&0.40	\\
\hline
\hline
\end{tabular}
\label{table1}
\end{table}

\section{Example of Building Grain Boundary}
In this section, we take the $\Sigma$5[001] grain boundary as an example to illustrate the procedure of generating CSL matrix and building grain boundary. 
For $\Sigma$ = 5 and [\textit{uvw}]=[001] , by using eqs. \ref{eqsigma} $-$ \ref{eq3}, we can get ($m$, $n$) = (1, 2), (1, 3), (2, 1) and (3, 1). By plugging these numbers into eq. \ref{eqtheta}, one can get $\theta$ = 126.9$^{\circ}$, 143.1$^{\circ}$, 53.1$^{\circ}$ and 36.9$^{\circ}$, respectively. As discussed before, ($m$, $n$) = (1,  2) and  ($m$, $n$) = (2, 1) (or $\theta$ = 126.9$^{\circ}$ and  $\theta$ = 53.1$^{\circ}$) refer to the same grain boundary, and also do for  ($m$, $n$) = (1,  3) and  ($m$, $n$) = (3, 1) (or $\theta$ = 143.1$^{\circ}$ and  $\theta$ = 36.9$^{\circ}$). 

 For the rotation axis [001], we have the unit vector \textbf{\textit{k}} = [0, 0, 1]. 
 Then using $\theta$ = 53.1$^{\circ}$ and eqs. \ref{eq5} $-$  \ref{eq6}, we are able to get a rotation matrix \textbf{R}:
 
\begin{gather*}
 \textbf{R}
 =  \frac{1}{5}
  \begin{bmatrix}
   3 & -4 & 0 \\
   4 & 3 & 0 \\
   0 & 0 & 5
   \end{bmatrix}
\end{gather*}
where $S$ = 5. 

Choose \textbf{U}:
\begin{gather*}
 \textbf{U}
 =
  \begin{bmatrix}
   1 & 0 & 1 \\
   0 & 1 & 0 \\
   0 & 1 & 1
   \end{bmatrix}
\end{gather*}
and plug it with \textbf{R} into eqs. \ref{eq13} and \ref{eq15}, we can get:
\begin{gather*}
 \textbf{T}^{\prime}
 = \frac{1}{5}
  \begin{bmatrix}
   2 & -4 & -5 \\
   4 & 2 & 0 \\
   4 & -3 & 0
   \end{bmatrix}
   \end{gather*}
and
   \begin{gather*}
{\textbf{X}^{\prime}}^{(0)}
 =
  \begin{bmatrix}
   0 & 3/4 & 1/2 \\
   0 & 1 & -1 \\
   -1 & -1/2 & 1
   \end{bmatrix}
\end{gather*}

\begin{figure}
\centering
\includegraphics[width=0.5\textwidth,clip]{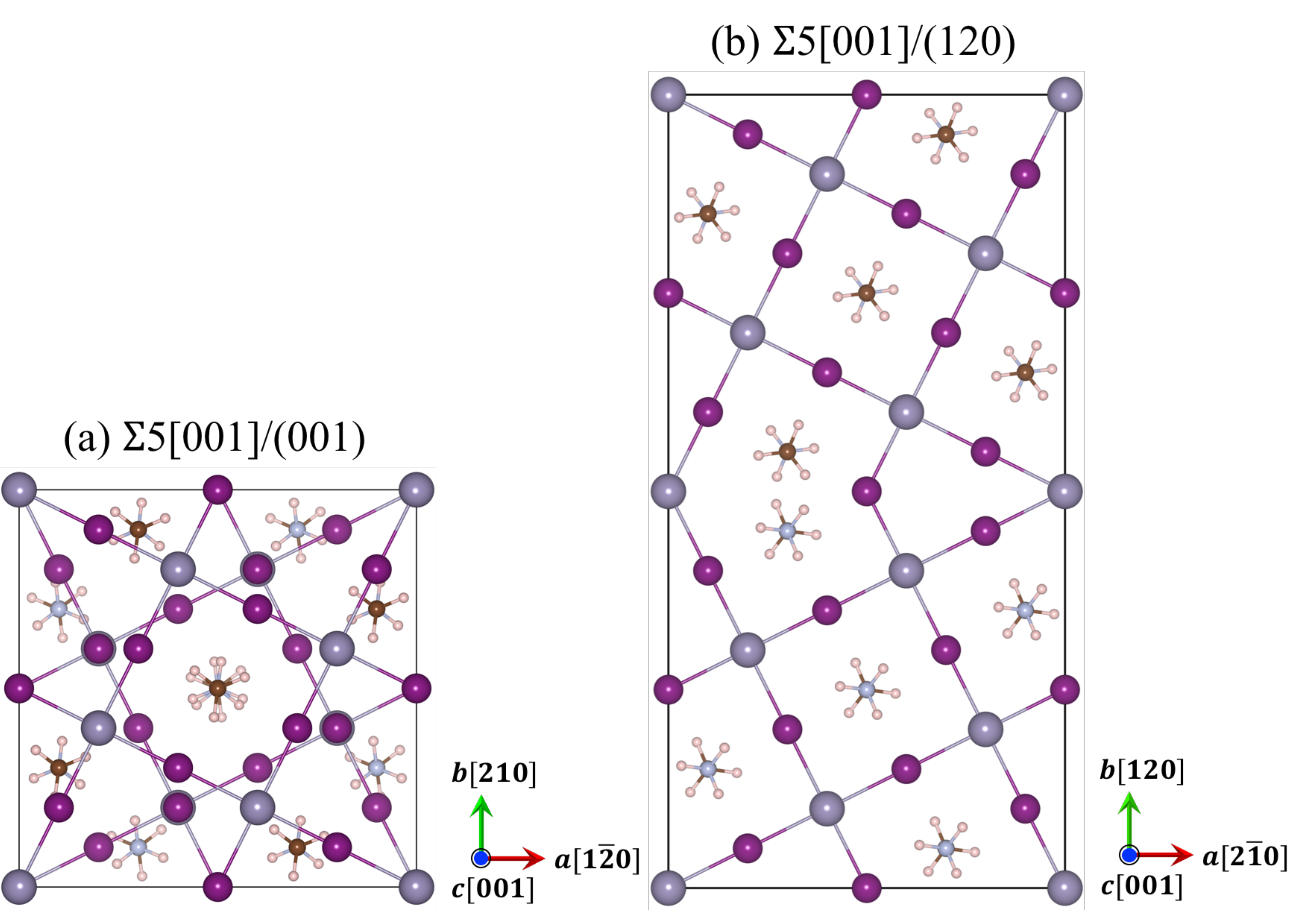}
\caption{Constructed grain boundary of cubic hybrid perovskite CH$_3$NH$_3$SnI$_3$: (a) $\Sigma$5[001]/(001) and (b) $\Sigma$5[001]/(120).}
\label{mapbi3_5_001}
\end{figure}

To obtain a CSL matrix, we make the matrix ${\textbf{X}^{\prime}}^{(0)}$ integral with each column vector having the shortest length, and meanwhile make its determinant equal to $\Sigma$, and get the following matrix \textbf{C}:
\begin{gather*}
 \textbf{C}
 =
  \begin{bmatrix}
   0 & -1 & 2 \\
   0 & 2 & 1 \\
   -1 & 0 & 0
   \end{bmatrix}
\end{gather*}
We further reshape the matrix \textbf{C} by making its each column vector orthogonal and setting its third column vector same with the rotation axis \textbf{\textit{o}}, and get the CSL matrix:
\begin{gather*}
 \textbf{C$^{\prime}$}
 =
  \begin{bmatrix}
   2 & 1 & 0 \\
   -1 & 2 & 0 \\
   0 & 0 & 1
   \end{bmatrix}
\end{gather*}
As mentioned above, each column vector of $\textbf{C$^{\prime}$}$ represents a grain boundary plane for $\Sigma$5[001] GB structure. Therefore, we have three grain boundary planes:  (2$\bar{1}$0), (120) and (001). 
$\Sigma$5[001]/(2$\bar{1}$0)  and $\Sigma$5[001]/(120) refer to the same tilt grain boundary, and $\Sigma$5[001]/(001) is a twist grain boundary. 
As a example, Fig. \ref{mapbi3_5_001} shows the built grain boundaries for cubic hybrid perovskite CH$_3$NH$_3$SnI$_3$: (a) $\Sigma$5[001]/(001) and (b) $\Sigma$5[001]/(120).\cite{Bernal_2014_jpcc}
 
\section{Conclusion}
We describe an algorithm implemented in an open-source python library for generating periodic tilt and twist grain boundary models in a universal fashion for \textit{ab-initio} and classical materials modeling.  The software framework first calculates the rotation angle from a given Sigma ($\Sigma$) value and rotation axis, and then generate a rotation matrix from the rotation angle. Next, the software will build CSL matrix using the rotation matrix and $\Sigma$ value. Finally, the software will generate two CSL grains from CSL matrix, and combine them together to build a grain boundary based on the selected grain boundary plane. This software framework is expected to enable high-throughput computational studies and materials informatics of grain boundaries in materials science. The source code is free  and available online: aimsgb.org.

\section{Acknowledgment}
This work was supported by the National Science Foundation under award number ACI-1550404. JL acknowledges the Vannevar Bush Faculty Fellowship program sponsored by the Basic Research Office of the Assistant Secretary of Defense for Research and Engineering (under the Office of Naval Research grant N00014-16-1-2569). This work used the Extreme Science and Engineering Discovery Environment (XSEDE), which is supported by National Science Foundation grant number OCI-1053575.
The authors thank Richard Tran, Hui Zheng, and Dr. Shyue Ping Ong for useful discussions.

\section{Appendix}
\begin{table}[th]
\small\addtolength{\tabcolsep}{4.8pt}
\renewcommand\thetable{A1}
\caption{The total required time (second) to generate grain boundary structures for SrTiO$_3$, cubic CH$_3$NH$_3$SnI$_3$, and tetragonal anatase TiO$_2$, including the time of writing to file. 
The total number (No.) of atoms in the grain boundary structure is listed for each $\Sigma$.}
\begin{tabular}{c|c|c|c|c|c|c}
\hline
\hline
& \multicolumn{2}{c|}{SrTiO$_3$}& \multicolumn{2}{c|}{CH$_3$NH$_3$SnI$_3$}& \multicolumn{2}{c}{TiO$_2$}\\
\hline
\multirow{2}{*}{$\Sigma$} & No. of & Time & No. of & Time & No. of & Time \\
& atoms & (sec) & atoms & (sec) & atoms & (sec)\\
\hhline{-------}
5	&50	&0.06	&120	&0.08	&120	&0.07	\\
13	&130	&0.03	&312	&0.10	&312	&0.10	\\
17	&170	&0.04	&408	&0.15	&408	&0.15	\\
25	&250	&0.07	&600	&0.30	&600	&0.29	\\
29	&290	&0.09	&696	&0.37	&696	&0.37	\\
37	&370	&0.13	&888	&0.57	&888	&0.57	\\
41	&410	&0.16	&984	&0.69	&984	&0.68	\\
53	&530	&0.23	&1272	&1.10	&1272	&1.12	\\
61	&610	&0.29	&1464	&1.43	&1464	&1.43	\\
65	&650	&0.32	&1560	&1.61	&1560	&1.59	\\
73	&730	&0.41	&1752	&2.01	&1752	&1.99	\\
85	&850	&0.52	&2040	&2.69	&2040	&2.68	\\
89	&890	&0.57	&2136	&2.91	&2136	&2.90	\\
97	&970	&0.68	&2328	&3.49	&2328	&3.46	\\
101	&1010	&0.71	&2424	&3.72	&2424	&3.69	\\
109	&1090	&0.84	&2616	&4.33	&2616	&4.34	\\
113	&1130	&0.88	&2712	&4.64	&2712	&4.60	\\
125	&1250	&1.07	&3000	&5.64	&3000	&5.67	\\
137	&1370	&1.29	&3288	&6.98	&3288	&6.74	\\
145	&1450	&1.40	&3480	&7.69	&3480	&7.45	\\
149	&1490	&1.48	&3576	&8.23	&3576	&8.01	\\
\hline
\hline
\end{tabular}
\label{A1}
\end{table}

\begin{table*}	
\renewcommand\thetable{A2}	
\small\addtolength{\tabcolsep}{2.8pt}
\caption{Lists of $\Sigma$, rotation angle $\theta$, GB plane and CSL matrix for rotation axis \textbf{\textit{o}} along [001], [110] and [111]. *Indicates the twisted grain boundary and the others are tilted grain boundary.}
\begin{tabular}{c|c|c|c|c|c|c|c|c|c|c|c}		
\hline		
\hline		
\multicolumn{4}{c|}{[001]}& \multicolumn{4}{c|}{[110]}& \multicolumn{4}{c}{[111]}\\		
\hline		
\multirow{2}{*}{$\Sigma$} &\multirow{2}{*}{$\theta$} &GB &\multirow{2}{*}{CSL}&\multirow{2}{*}{$\Sigma$} &\multirow{2}{*}{$\theta$} &GB &\multirow{2}{*}{CSL}&\multirow{2}{*}{$\Sigma$} &\multirow{2}{*}{$\theta$} &GB &\multirow{2}{*}{CSL}\\		
& &Plane & & & &Plane & & & &Plane & \\		
\hline		
\multirow{4}{*}{5}&53.13&$\begin{array}{ccc}(2\bar{1}0)\\(120)\\(001)*\end{array}$&$\begin{pmatrix}2&1&0\\-1&2&0\\0&0&1\end{pmatrix}$ &	\multirow{4}{*}{3}&\multirow{4}{*}{70.53}&\multirow{4}{*}{$\begin{array}{ccc}(\bar{1}11)\\(1\bar{1}2)\\(110)*\end{array}$}&\multirow{4}{*}{$\begin{pmatrix}-1&1&1\\1&-1&1\\1&2&0\end{pmatrix}$} &	\multirow{4}{*}{3}&\multirow{4}{*}{60.00}&\multirow{4}{*}{$\begin{array}{ccc}(1\bar{2}1)\\(10\bar{1})\\(111)*\end{array}$}&\multirow{4}{*}{$\begin{pmatrix}1&1&1\\-2&0&1\\1&-1&1\end{pmatrix}$} \\
\hhline{~---~~~~~~~~}		
&36.87&$\begin{array}{ccc}(3\bar{1}0)\\(130)\\(001)*\end{array}$&$\begin{pmatrix}3&1&0\\-1&3&0\\0&0&1\end{pmatrix}$ &	&&&	&&& \\
\hline		
\multirow{4}{*}{13}&67.38&$\begin{array}{ccc}(3\bar{2}0)\\(230)\\(001)*\end{array}$&$\begin{pmatrix}3&2&0\\-2&3&0\\0&0&1\end{pmatrix}$ &	\multirow{4}{*}{9}&\multirow{4}{*}{38.94}&\multirow{4}{*}{$\begin{array}{ccc}(\bar{1}1\bar{4})\\(\bar{2}21)\\(110)*\end{array}$}&\multirow{4}{*}{$\begin{pmatrix}-1&-2&1\\1&2&1\\-4&1&0\end{pmatrix}$} &	\multirow{4}{*}{7}&\multirow{4}{*}{38.21}&\multirow{4}{*}{$\begin{array}{ccc}(1\bar{3}2)\\(5\bar{1}\bar{4})\\(111)*\end{array}$}&\multirow{4}{*}{$\begin{pmatrix}1&5&1\\-3&-1&1\\2&-4&1\end{pmatrix}$} \\
\hhline{~---~~~~~~~~}		
&22.62&$\begin{array}{ccc}(5\bar{1}0)\\(150)\\(001)*\end{array}$&$\begin{pmatrix}5&1&0\\-1&5&0\\0&0&1\end{pmatrix}$ &	&&&	&&& \\
\hline		
\multirow{4}{*}{17}&28.07&$\begin{array}{ccc}(410)\\(\bar{1}40)\\(001)*\end{array}$&$\begin{pmatrix}4&-1&0\\1&4&0\\0&0&1\end{pmatrix}$ &	\multirow{4}{*}{11}&\multirow{4}{*}{50.48}&\multirow{4}{*}{$\begin{array}{ccc}(\bar{3}32)\\(1\bar{1}3)\\(110)*\end{array}$}&\multirow{4}{*}{$\begin{pmatrix}-3&1&1\\3&-1&1\\2&3&0\end{pmatrix}$} &	\multirow{4}{*}{13}&\multirow{4}{*}{27.80}&\multirow{4}{*}{$\begin{array}{ccc}(1\bar{4}3)\\(7\bar{2}\bar{5})\\(111)*\end{array}$}&\multirow{4}{*}{$\begin{pmatrix}1&7&1\\-4&-2&1\\3&-5&1\end{pmatrix}$} \\
\hhline{~---~~~~~~~~}		
&61.93&$\begin{array}{ccc}(530)\\(\bar{3}50)\\(001)*\end{array}$&$\begin{pmatrix}5&-3&0\\3&5&0\\0&0&1\end{pmatrix}$ &	&&&	&&& \\
\hline		
\multirow{4}{*}{25}&73.74&$\begin{array}{ccc}(4\bar{3}0)\\(340)\\(001)*\end{array}$&$\begin{pmatrix}4&3&0\\-3&4&0\\0&0&1\end{pmatrix}$ &	\multirow{4}{*}{17}&\multirow{4}{*}{86.63}&\multirow{4}{*}{$\begin{array}{ccc}(2\bar{2}3)\\(3\bar{3}\bar{4})\\(110)*\end{array}$}&\multirow{4}{*}{$\begin{pmatrix}2&3&1\\-2&-3&1\\3&-4&0\end{pmatrix}$} &	\multirow{4}{*}{19}&\multirow{4}{*}{46.83}&\multirow{4}{*}{$\begin{array}{ccc}(2\bar{5}3)\\(8\bar{1}\bar{7})\\(111)*\end{array}$}&\multirow{4}{*}{$\begin{pmatrix}2&8&1\\-5&-1&1\\3&-7&1\end{pmatrix}$} \\
\hhline{~---~~~~~~~~}		
&16.26&$\begin{array}{ccc}(7\bar{1}0)\\(170)\\(001)*\end{array}$&$\begin{pmatrix}7&1&0\\-1&7&0\\0&0&1\end{pmatrix}$ &	&&&	&&& \\
\hline		
\multirow{4}{*}{29}&43.60&$\begin{array}{ccc}(\bar{2}50)\\(\bar{5}\bar{2}0)\\(001)*\end{array}$&$\begin{pmatrix}-2&-5&0\\5&-2&0\\0&0&1\end{pmatrix}$ &	\multirow{4}{*}{19}&\multirow{4}{*}{26.53}&\multirow{4}{*}{$\begin{array}{ccc}(\bar{3}31)\\(1\bar{1}6)\\(110)*\end{array}$}&\multirow{4}{*}{$\begin{pmatrix}-3&1&1\\3&-1&1\\1&6&0\end{pmatrix}$} &	\multirow{4}{*}{21 (7)}&\multirow{4}{*}{21.79}&\multirow{4}{*}{$\begin{array}{ccc}(1\bar{5}4)\\(3\bar{1}\bar{2})\\(111)*\end{array}$}&\multirow{4}{*}{$\begin{pmatrix}1&3&1\\-5&-1&1\\4&-2&1\end{pmatrix}$} \\
\hhline{~---~~~~~~~~}		
&46.40&$\begin{array}{ccc}(\bar{3}70)\\(\bar{7}\bar{3}0)\\(001)*\end{array}$&$\begin{pmatrix}-3&-7&0\\7&-3&0\\0&0&1\end{pmatrix}$ &	&&&	&&& \\
\hline		
\multirow{4}{*}{37}&18.92&$\begin{array}{ccc}(610)\\(\bar{1}60)\\(001)*\end{array}$&$\begin{pmatrix}6&-1&0\\1&6&0\\0&0&1\end{pmatrix}$ &	\multirow{4}{*}{27}&\multirow{4}{*}{31.59}&\multirow{4}{*}{$\begin{array}{ccc}(\bar{5}52)\\(1\bar{1}5)\\(110)*\end{array}$}&\multirow{4}{*}{$\begin{pmatrix}-5&1&1\\5&-1&1\\2&5&0\end{pmatrix}$} &	\multirow{4}{*}{31}&\multirow{4}{*}{17.90}&\multirow{4}{*}{$\begin{array}{ccc}(1&\bar{6}&5)\\(11&\bar{4}&\bar{7})\\(1&1&1)*\end{array}$}&\multirow{4}{*}{$\begin{pmatrix}1&11&1\\-6&-4&1\\5&-7&1\end{pmatrix}$} \\
\hhline{~---~~~~~~~~}		
&71.08&$\begin{array}{ccc}(750)\\(\bar{5}70)\\(001)*\end{array}$&$\begin{pmatrix}7&-5&0\\5&7&0\\0&0&1\end{pmatrix}$ &	&&&	&&& \\
\hline		
\multirow{4}{*}{41}&77.32&$\begin{array}{ccc}(5\bar{4}0)\\(450)\\(001)*\end{array}$&$\begin{pmatrix}5&4&0\\-4&5&0\\0&0&1\end{pmatrix}$ &	\multirow{4}{*}{33}&\multirow{4}{*}{20.05}&\multirow{4}{*}{$\begin{array}{ccc}(\bar{4}41)\\(1\bar{1}8)\\(110)*\end{array}$}&\multirow{4}{*}{$\begin{pmatrix}-4&1&1\\4&-1&1\\1&8&0\end{pmatrix}$} &	\multirow{4}{*}{37}&\multirow{4}{*}{50.57}&\multirow{4}{*}{$\begin{array}{ccc}(3&\bar{7}&4)\\(11&\bar{1}&\bar{10})\\(1&1&1)*\end{array}$}&\multirow{4}{*}{$\begin{pmatrix}3&11&1\\-7&-1&1\\4&-10&1\end{pmatrix}$} \\
\hhline{~---~~~~~~~~}		
&12.68&$\begin{array}{ccc}(9\bar{1}0)\\(190)\\(001)*\end{array}$&$\begin{pmatrix}9&1&0\\-1&9&0\\0&0&1\end{pmatrix}$ &	&&&	&&& \\
\hline		
\multirow{4}{*}{53}&31.89&$\begin{array}{ccc}(\bar{2}70)\\(\bar{7}\bar{2}0)\\(001)*\end{array}$&$\begin{pmatrix}-2&-7&0\\7&-2&0\\0&0&1\end{pmatrix}$ &	\multirow{4}{*}{41}&\multirow{4}{*}{55.88}&\multirow{4}{*}{$\begin{array}{ccc}(\bar{3}3\bar{8})\\(\bar{4}43)\\(110)*\end{array}$}&\multirow{4}{*}{$\begin{pmatrix}-3&-4&1\\3&4&1\\-8&3&0\end{pmatrix}$} &	\multirow{4}{*}{39 (13)}&\multirow{4}{*}{32.20}&\multirow{4}{*}{$\begin{array}{ccc}(2\bar{7}5)\\(4\bar{1}\bar{3})\\(111)*\end{array}$}&\multirow{4}{*}{$\begin{pmatrix}2&4&1\\-7&-1&1\\5&-3&1\end{pmatrix}$} \\
\hhline{~---~~~~~~~~}		
&58.11&$\begin{array}{ccc}(\bar{5}90)\\(\bar{9}\bar{5}0)\\(001)*\end{array}$&$\begin{pmatrix}-5&-9&0\\9&-5&0\\0&0&1\end{pmatrix}$ &	&&&	&&& \\
\hline		
\hline		
\end{tabular}		
\end{table*}		
		
\begin{table*}		
\begin{tabular}{c|c|c|c|c|c|c|c|c|c|c|c}		
\hline		
\hline		
\multicolumn{4}{c|}{[001]}& \multicolumn{4}{c|}{[110]}& \multicolumn{4}{c}{[111]}\\		
\hline		
\multirow{2}{*}{$\Sigma$} &\multirow{2}{*}{$\theta$} &GB &\multirow{2}{*}{CSL}&\multirow{2}{*}{$\Sigma$} &\multirow{2}{*}{$\theta$} &GB &\multirow{2}{*}{CSL}&\multirow{2}{*}{$\Sigma$} &\multirow{2}{*}{$\theta$} &GB &\multirow{2}{*}{CSL}\\		
& &Plane & & & &Plane & & & &Plane & \\		
\hline		
\multirow{4}{*}{61}&79.61&$\begin{array}{ccc}(6\bar{5}0)\\(560)\\(001)*\end{array}$&$\begin{pmatrix}6&5&0\\-5&6&0\\0&0&1\end{pmatrix}$ &	\multirow{4}{*}{43}&\multirow{4}{*}{80.63}&\multirow{4}{*}{$\begin{array}{ccc}(3\bar{3}5)\\(5\bar{5}\bar{6})\\(110)*\end{array}$}&\multirow{4}{*}{$\begin{pmatrix}3&5&1\\-3&-5&1\\5&-6&0\end{pmatrix}$} &	\multirow{4}{*}{43}&\multirow{4}{*}{15.18}&\multirow{4}{*}{$\begin{array}{ccc}(1&\bar{7}&6)\\(13&\bar{5}&\bar{8})\\(1&1&1)*\end{array}$}&\multirow{4}{*}{$\begin{pmatrix}1&13&1\\-7&-5&1\\6&-8&1\end{pmatrix}$} \\
\hhline{~---~~~~~~~~}		
&10.39&$\begin{array}{ccc}(11&\bar{1}&0)\\(1&11&0)\\(0&0&1)*\end{array}$&$\begin{pmatrix}11&1&0\\-1&11&0\\0&0&1\end{pmatrix}$ &	&&&	&&& \\
\hline		
\multirow{4}{*}{65}&14.25&$\begin{array}{ccc}(810)\\(\bar{1}80)\\(001)*\end{array}$&$\begin{pmatrix}8&-1&0\\1&8&0\\0&0&1\end{pmatrix}$ &	\multirow{4}{*}{51}&\multirow{4}{*}{16.10}&\multirow{4}{*}{$\begin{array}{ccc}(\bar{5}&5&1)\\(1&\bar{1}&10)\\(1&1&0)*\end{array}$}&\multirow{4}{*}{$\begin{pmatrix}-5&1&1\\5&-1&1\\1&10&0\end{pmatrix}$} &	\multirow{4}{*}{49}&\multirow{4}{*}{43.57}&\multirow{4}{*}{$\begin{array}{ccc}(3&\bar{8}&5)\\(13&\bar{2}&\bar{11})\\(1&1&1)*\end{array}$}&\multirow{4}{*}{$\begin{pmatrix}3&13&1\\-8&-2&1\\5&-11&1\end{pmatrix}$} \\
\hhline{~---~~~~~~~~}		
&75.75&$\begin{array}{ccc}(11&3&0)\\(\bar{3}&11&0)\\(0&0&1)*\end{array}$&$\begin{pmatrix}11&-3&0\\3&11&0\\0&0&1\end{pmatrix}$ &	&&&	&&& \\
\hline		
\multirow{4}{*}{73}&41.11&$\begin{array}{ccc}(\bar{3}80)\\(\bar{8}\bar{3}0)\\(001)*\end{array}$&$\begin{pmatrix}-3&-8&0\\8&-3&0\\0&0&1\end{pmatrix}$ &	\multirow{4}{*}{57}&\multirow{4}{*}{44.00}&\multirow{4}{*}{$\begin{array}{ccc}(\bar{7}74)\\(2\bar{2}7)\\(110)*\end{array}$}&\multirow{4}{*}{$\begin{pmatrix}-7&2&1\\7&-2&1\\4&7&0\end{pmatrix}$} &	\multirow{4}{*}{57 (19)}&\multirow{4}{*}{13.17}&\multirow{4}{*}{$\begin{array}{ccc}(1\bar{8}7)\\(5\bar{2}\bar{3})\\(111)*\end{array}$}&\multirow{4}{*}{$\begin{pmatrix}1&5&1\\-8&-2&1\\7&-3&1\end{pmatrix}$} \\
\hhline{~---~~~~~~~~}		
&48.89&$\begin{array}{ccc}(\bar{5}&11&0)\\(\bar{11}&\bar{5}&0)\\(0&0&1)*\end{array}$&$\begin{pmatrix}-5&-11&0\\11&-5&0\\0&0&1\end{pmatrix}$ &	&&&	&&& \\
\hline		
\multirow{4}{*}{85}&81.20&$\begin{array}{ccc}(7\bar{6}0)\\(670)\\(001)*\end{array}$&$\begin{pmatrix}7&6&0\\-6&7&0\\0&0&1\end{pmatrix}$ &	\multirow{4}{*}{59}&\multirow{4}{*}{45.98}&\multirow{4}{*}{$\begin{array}{ccc}(\bar{3}&3&\bar{10})\\(\bar{5}&5&3)\\(1&1&0)*\end{array}$}&\multirow{4}{*}{$\begin{pmatrix}-3&-5&1\\3&5&1\\-10&3&0\end{pmatrix}$} &	\multirow{4}{*}{61}&\multirow{4}{*}{52.66}&\multirow{4}{*}{$\begin{array}{ccc}(4&\bar{9}&5)\\(14&\bar{1}&\bar{13})\\(1&1&1)*\end{array}$}&\multirow{4}{*}{$\begin{pmatrix}4&14&1\\-9&-1&1\\5&-13&1\end{pmatrix}$} \\
\hhline{~---~~~~~~~~}		
&8.80&$\begin{array}{ccc}(13&\bar{1}&0)\\(1&13&0)\\(0&0&1)*\end{array}$&$\begin{pmatrix}13&1&0\\-1&13&0\\0&0&1\end{pmatrix}$ &	&&&	&&& \\
\hline		
\multirow{4}{*}{89}&64.01&$\begin{array}{ccc}(8\bar{5}0)\\(580)\\(001)*\end{array}$&$\begin{pmatrix}8&5&0\\-5&8&0\\0&0&1\end{pmatrix}$ &	\multirow{4}{*}{67}&\multirow{4}{*}{62.44}&\multirow{4}{*}{$\begin{array}{ccc}(\bar{7}76)\\(3\bar{3}7)\\(110)*\end{array}$}&\multirow{4}{*}{$\begin{pmatrix}-7&3&1\\7&-3&1\\6&7&0\end{pmatrix}$} &	\multirow{4}{*}{67}&\multirow{4}{*}{24.43}&\multirow{4}{*}{$\begin{array}{ccc}(2&\bar{9}&7)\\(16&\bar{5}&\bar{11})\\(1&1&1)*\end{array}$}&\multirow{4}{*}{$\begin{pmatrix}2&16&1\\-9&-5&1\\7&-11&1\end{pmatrix}$} \\
\hhline{~---~~~~~~~~}		
&25.99&$\begin{array}{ccc}(13&\bar{3}&0)\\(3&13&0)\\(0&0&1)*\end{array}$&$\begin{pmatrix}13&3&0\\-3&13&0\\0&0&1\end{pmatrix}$ &	&&&	&&& \\
\hline		
\multirow{4}{*}{97}&47.92&$\begin{array}{ccc}(9\bar{4}0)\\(490)\\(001)*\end{array}$&$\begin{pmatrix}9&4&0\\-4&9&0\\0&0&1\end{pmatrix}$ &	\multirow{4}{*}{73}&\multirow{4}{*}{13.44}&\multirow{4}{*}{$\begin{array}{ccc}(\bar{6}&6&1)\\(1&\bar{1}&12)\\(1&1&0)*\end{array}$}&\multirow{4}{*}{$\begin{pmatrix}-6&1&1\\6&-1&1\\1&12&0\end{pmatrix}$} &	\multirow{4}{*}{73}&\multirow{4}{*}{11.64}&\multirow{4}{*}{$\begin{array}{ccc}(1&\bar{9}&8)\\(17&\bar{7}&\bar{10})\\(1&1&1)*\end{array}$}&\multirow{4}{*}{$\begin{pmatrix}1&17&1\\-9&-7&1\\8&-10&1\end{pmatrix}$} \\
\hhline{~---~~~~~~~~}		
&42.08&$\begin{array}{ccc}(13&\bar{5}&0)\\(5&13&0)\\(0&0&1)*\end{array}$&$\begin{pmatrix}13&5&0\\-5&13&0\\0&0&1\end{pmatrix}$ &	&&&	&&& \\
\hline		
\multirow{4}{*}{101}&11.42&$\begin{array}{ccc}(10&1&0)\\(\bar{1}&10&0)\\(0&0&1)*\end{array}$&$\begin{pmatrix}10&-1&0\\1&10&0\\0&0&1\end{pmatrix}$ &	\multirow{4}{*}{81}&\multirow{4}{*}{77.88}&\multirow{4}{*}{$\begin{array}{ccc}(4\bar{4}7)\\(7\bar{7}\bar{8})\\(110)*\end{array}$}&\multirow{4}{*}{$\begin{pmatrix}4&7&1\\-4&-7&1\\7&-8&0\end{pmatrix}$} &	\multirow{4}{*}{79}&\multirow{4}{*}{33.99}&\multirow{4}{*}{$\begin{array}{ccc}(3&\bar{10}&7)\\(17&\bar{4}&\bar{13})\\(1&1&1)*\end{array}$}&\multirow{4}{*}{$\begin{pmatrix}3&17&1\\-10&-4&1\\7&-13&1\end{pmatrix}$} \\
\hhline{~---~~~~~~~~}		
&78.58&$\begin{array}{ccc}(11&9&0)\\(\bar{9}&11&0)\\(0&0&1)*\end{array}$&$\begin{pmatrix}11&-9&0\\9&11&0\\0&0&1\end{pmatrix}$ &	&&&	&&& \\
\hline		
\multirow{4}{*}{109}&33.40&$\begin{array}{ccc}(\bar{3}&10&0)\\(\bar{10}&\bar{3}&0)\\(0&0&1)*\end{array}$&$\begin{pmatrix}-3&-10&0\\10&-3&0\\0&0&1\end{pmatrix}$ &	\multirow{4}{*}{83}&\multirow{4}{*}{17.86}&\multirow{4}{*}{$\begin{array}{ccc}(\bar{9}92)\\(1\bar{1}9)\\(110)*\end{array}$}&\multirow{4}{*}{$\begin{pmatrix}-9&1&1\\9&-1&1\\2&9&0\end{pmatrix}$} &	\multirow{4}{*}{91}&\multirow{4}{*}{10.42}&\multirow{4}{*}{$\begin{array}{ccc}(1&\bar{10}&9)\\(19&\bar{8}&\bar{11})\\(1&1&1)*\end{array}$}&\multirow{4}{*}{$\begin{pmatrix}1&19&1\\-10&-8&1\\9&-11&1\end{pmatrix}$} \\
\hhline{~---~~~~~~~~}		
&56.60&$\begin{array}{ccc}(\bar{7}&13&0)\\(\bar{13}&\bar{7}&0)\\(0&0&1)*\end{array}$&$\begin{pmatrix}-7&-13&0\\13&-7&0\\0&0&1\end{pmatrix}$ &	&&&	&&& \\
\hline		
\hline		
\end{tabular}		
\end{table*}		
		
\begin{table*}		
\begin{tabular}{c|c|c|c|c|c|c|c|c|c|c|c}		
\hline		
\hline		
\multicolumn{4}{c|}{[001]}& \multicolumn{4}{c|}{[110]}& \multicolumn{4}{c}{[111]}\\		
\hline		
\multirow{2}{*}{$\Sigma$} &\multirow{2}{*}{$\theta$} &GB &\multirow{2}{*}{CSL}&\multirow{2}{*}{$\Sigma$} &\multirow{2}{*}{$\theta$} &GB &\multirow{2}{*}{CSL}&\multirow{2}{*}{$\Sigma$} &\multirow{2}{*}{$\theta$} &GB &\multirow{2}{*}{CSL}\\		
& &Plane & & & &Plane & & & &Plane & \\		
\hline		
\multirow{4}{*}{113}&82.37&$\begin{array}{ccc}(8\bar{7}0)\\(780)\\(001)*\end{array}$&$\begin{pmatrix}8&7&0\\-7&8&0\\0&0&1\end{pmatrix}$ &	\multirow{4}{*}{89}&\multirow{4}{*}{34.89}&\multirow{4}{*}{$\begin{array}{ccc}(\bar{9}94)\\(2\bar{2}9)\\(110)*\end{array}$}&\multirow{4}{*}{$\begin{pmatrix}-9&2&1\\9&-2&1\\4&9&0\end{pmatrix}$} &	\multirow{4}{*}{93 (31)}&\multirow{4}{*}{42.10}&\multirow{4}{*}{$\begin{array}{ccc}(4&\bar{11}&7)\\(6&\bar{1}&\bar{5})\\(1&1&1)*\end{array}$}&\multirow{4}{*}{$\begin{pmatrix}4&6&1\\-11&-1&1\\7&-5&1\end{pmatrix}$} \\
\hhline{~---~~~~~~~~}		
&7.63&$\begin{array}{ccc}(15&\bar{1}&0)\\(1&15&0)\\(0&0&1)*\end{array}$&$\begin{pmatrix}15&1&0\\-1&15&0\\0&0&1\end{pmatrix}$ &	&&&	&&& \\
\hline		
\multirow{4}{*}{125}&20.61&$\begin{array}{ccc}(\bar{2}&11&0)\\(\bar{11}&\bar{2}&0)\\(0&0&1)*\end{array}$&$\begin{pmatrix}-2&-11&0\\11&-2&0\\0&0&1\end{pmatrix}$ &	\multirow{4}{*}{97}&\multirow{4}{*}{61.02}&\multirow{4}{*}{$\begin{array}{ccc}(\bar{5}&5&\bar{12})\\(\bar{6}&6&5)\\(1&1&0)*\end{array}$}&\multirow{4}{*}{$\begin{pmatrix}-5&-6&1\\5&6&1\\-12&5&0\end{pmatrix}$} &	\multirow{4}{*}{97}&\multirow{4}{*}{30.59}&\multirow{4}{*}{$\begin{array}{ccc}(3&\bar{11}&8)\\(19&\bar{5}&\bar{14})\\(1&1&1)*\end{array}$}&\multirow{4}{*}{$\begin{pmatrix}3&19&1\\-11&-5&1\\8&-14&1\end{pmatrix}$} \\
\hhline{~---~~~~~~~~}		
&69.39&$\begin{array}{ccc}(\bar{9}&13&0)\\(\bar{13}&\bar{9}&0)\\(0&0&1)*\end{array}$&$\begin{pmatrix}-9&-13&0\\13&-9&0\\0&0&1\end{pmatrix}$ &	&&&	&&& \\
\hline		
\multirow{4}{*}{137}&39.97&$\begin{array}{ccc}(\bar{4}&11&0)\\(\bar{11}&\bar{4}&0)\\(0&0&1)*\end{array}$&$\begin{pmatrix}-4&-11&0\\11&-4&0\\0&0&1\end{pmatrix}$ &	\multirow{4}{*}{99}&\multirow{4}{*}{11.54}&\multirow{4}{*}{$\begin{array}{ccc}(\bar{7}&7&1)\\(1&\bar{1}&14)\\(1&1&0)*\end{array}$}&\multirow{4}{*}{$\begin{pmatrix}-7&1&1\\7&-1&1\\1&14&0\end{pmatrix}$} &	\multirow{4}{*}{103}&\multirow{4}{*}{19.65}&\multirow{4}{*}{$\begin{array}{ccc}(2&\bar{11}&9)\\(20&\bar{7}&\bar{13})\\(1&1&1)*\end{array}$}&\multirow{4}{*}{$\begin{pmatrix}2&20&1\\-11&-7&1\\9&-13&1\end{pmatrix}$} \\
\hhline{~---~~~~~~~~}		
&50.03&$\begin{array}{ccc}(\bar{7}&15&0)\\(\bar{15}&\bar{7}&0)\\(0&0&1)*\end{array}$&$\begin{pmatrix}-7&-15&0\\15&-7&0\\0&0&1\end{pmatrix}$ &	&&&	&&& \\
\hline		
\multirow{4}{*}{145}&83.27&$\begin{array}{ccc}(9\bar{8}0)\\(890)\\(001)*\end{array}$&$\begin{pmatrix}9&8&0\\-8&9&0\\0&0&1\end{pmatrix}$ &	\multirow{4}{*}{107}&\multirow{4}{*}{33.72}&\multirow{4}{*}{$\begin{array}{ccc}(\bar{3}&3&\bar{14})\\(\bar{7}&7&3)\\(1&1&0)*\end{array}$}&\multirow{4}{*}{$\begin{pmatrix}-3&-7&1\\3&7&1\\-14&3&0\end{pmatrix}$} &	\multirow{4}{*}{109}&\multirow{4}{*}{49.01}&\multirow{4}{*}{$\begin{array}{ccc}(5&\bar{12}&7)\\(19&\bar{2}&\bar{17})\\(1&1&1)*\end{array}$}&\multirow{4}{*}{$\begin{pmatrix}5&19&1\\-12&-2&1\\7&-17&1\end{pmatrix}$} \\
\hhline{~---~~~~~~~~}		
&6.73&$\begin{array}{ccc}(17&\bar{1}&0)\\(1&17&0)\\(0&0&1)*\end{array}$&$\begin{pmatrix}17&1&0\\-1&17&0\\0&0&1\end{pmatrix}$ &	&&&	&&& \\
\hline		
\multirow{4}{*}{149}&69.98&$\begin{array}{ccc}(\bar{10}&7&0)\\(\bar{7}&\bar{10}&0)\\(0&0&1)*\end{array}$&$\begin{pmatrix}-10&-7&0\\7&-10&0\\0&0&1\end{pmatrix}$ &	\multirow{4}{*}{113}&\multirow{4}{*}{64.30}&\multirow{4}{*}{$\begin{array}{ccc}(\bar{9}98)\\(4\bar{4}9)\\(110)*\end{array}$}&\multirow{4}{*}{$\begin{pmatrix}-9&4&1\\9&-4&1\\8&9&0\end{pmatrix}$} &	\multirow{4}{*}{111 (37)}&\multirow{4}{*}{9.43}&\multirow{4}{*}{$\begin{array}{ccc}(1&\bar{11}&10)\\(7&\bar{3}&\bar{4})\\(1&1&1)*\end{array}$}&\multirow{4}{*}{$\begin{pmatrix}1&7&1\\-11&-3&1\\10&-4&1\end{pmatrix}$} \\
\hhline{~---~~~~~~~~}		
&20.02&$\begin{array}{ccc}(\bar{17}&3&0)\\(\bar{3}&\bar{17}&0)\\(0&0&1)*\end{array}$&$\begin{pmatrix}-17&-3&0\\3&-17&0\\0&0&1\end{pmatrix}$ &	&&&	&&& \\
\hline		
\multirow{4}{*}{157}&57.22&$\begin{array}{ccc}(\bar{11}&6&0)\\(\bar{6}&\bar{11}&0)\\(0&0&1)*\end{array}$&$\begin{pmatrix}-11&-6&0\\6&-11&0\\0&0&1\end{pmatrix}$ &	\multirow{4}{*}{121}&\multirow{4}{*}{79.04}&\multirow{4}{*}{$\begin{array}{ccc}(7&\bar{7}&12)\\(6&\bar{6}&\bar{7})\\(1&1&0)*\end{array}$}&\multirow{4}{*}{$\begin{pmatrix}7&6&1\\-7&-6&1\\12&-7&0\end{pmatrix}$} &	\multirow{4}{*}{127}&\multirow{4}{*}{54.91}&\multirow{4}{*}{$\begin{array}{ccc}(6&\bar{13}&7)\\(20&\bar{1}&\bar{19})\\(1&1&1)*\end{array}$}&\multirow{4}{*}{$\begin{pmatrix}6&20&1\\-13&-1&1\\7&-19&1\end{pmatrix}$} \\
\hhline{~---~~~~~~~~}		
&32.78&$\begin{array}{ccc}(\bar{17}&5&0)\\(\bar{5}&\bar{17}&0)\\(0&0&1)*\end{array}$&$\begin{pmatrix}-17&-5&0\\5&-17&0\\0&0&1\end{pmatrix}$ &	&&&	&&& \\
\hline		
\multirow{4}{*}{169}&45.24&$\begin{array}{ccc}(\bar{12}&5&0)\\(\bar{5}&\bar{12}&0)\\(0&0&1)*\end{array}$&$\begin{pmatrix}-12&-5&0\\5&-12&0\\0&0&1\end{pmatrix}$ &	\multirow{4}{*}{123}&\multirow{4}{*}{14.65}&\multirow{4}{*}{$\begin{array}{ccc}(\bar{11}&11&2)\\(1&\bar{1}&11)\\(1&1&0)*\end{array}$}&\multirow{4}{*}{$\begin{pmatrix}-11&1&1\\11&-1&1\\2&11&0\end{pmatrix}$} &	\multirow{4}{*}{129 (43)}&\multirow{4}{*}{44.82}&\multirow{4}{*}{$\begin{array}{ccc}(5&\bar{13}&8)\\(7&\bar{1}&\bar{6})\\(1&1&1)*\end{array}$}&\multirow{4}{*}{$\begin{pmatrix}5&7&1\\-13&-1&1\\8&-6&1\end{pmatrix}$} \\
\hhline{~---~~~~~~~~}		
&44.76&$\begin{array}{ccc}(\bar{17}&7&0)\\(\bar{7}&\bar{17}&0)\\(0&0&1)*\end{array}$&$\begin{pmatrix}-17&-7&0\\7&-17&0\\0&0&1\end{pmatrix}$ &	&&&	&&& \\
\hline		
\multirow{4}{*}{173}&17.49&$\begin{array}{ccc}(\bar{2}&13&0)\\(\bar{13}&\bar{2}&0)\\(0&0&1)*\end{array}$&$\begin{pmatrix}-2&-13&0\\13&-2&0\\0&0&1\end{pmatrix}$ &	\multirow{4}{*}{129}&\multirow{4}{*}{10.10}&\multirow{4}{*}{$\begin{array}{ccc}(\bar{8}&8&1)\\(1&\bar{1}&16)\\(1&1&0)*\end{array}$}&\multirow{4}{*}{$\begin{pmatrix}-8&1&1\\8&-1&1\\1&16&0\end{pmatrix}$} &	\multirow{4}{*}{133}&\multirow{4}{*}{8.61}&\multirow{4}{*}{$\begin{array}{ccc}(1&\bar{12}&11)\\(23&\bar{10}&\bar{13})\\(1&1&1)*\end{array}$}&\multirow{4}{*}{$\begin{pmatrix}1&23&1\\-12&-10&1\\11&-13&1\end{pmatrix}$} \\
\hhline{~---~~~~~~~~}		
&72.51&$\begin{array}{ccc}(\bar{11}&15&0)\\(\bar{15}&\bar{11}&0)\\(0&0&1)*\end{array}$&$\begin{pmatrix}-11&-15&0\\15&-11&0\\0&0&1\end{pmatrix}$ &	&&&	&&& \\
\hline		
\hline		
\end{tabular}		
\end{table*}		
		
\begin{table*}		
\begin{tabular}{c|c|c|c|c|c|c|c|c|c|c|c}		
\hline		
\hline		
\multicolumn{4}{c|}{[001]}& \multicolumn{4}{c|}{[110]}& \multicolumn{4}{c}{[111]}\\		
\hline		
\multirow{2}{*}{$\Sigma$} &\multirow{2}{*}{$\theta$} &GB &\multirow{2}{*}{CSL}&\multirow{2}{*}{$\Sigma$} &\multirow{2}{*}{$\theta$} &GB &\multirow{2}{*}{CSL}&\multirow{2}{*}{$\Sigma$} &\multirow{2}{*}{$\theta$} &GB &\multirow{2}{*}{CSL}\\		
& &Plane & & & &Plane & & & &Plane & \\		
\hline		
\multirow{4}{*}{181}&83.97&$\begin{array}{ccc}(10&\bar{9}&0)\\(9&10&0)\\(0&0&1)*\end{array}$&$\begin{pmatrix}10&9&0\\-9&10&0\\0&0&1\end{pmatrix}$ &	\multirow{4}{*}{131}&\multirow{4}{*}{76.31}&\multirow{4}{*}{$\begin{array}{ccc}(5&\bar{5}&9)\\(9&\bar{9}&\bar{10})\\(1&1&0)*\end{array}$}&\multirow{4}{*}{$\begin{pmatrix}5&9&1\\-5&-9&1\\9&-10&0\end{pmatrix}$} &	\multirow{4}{*}{139}&\multirow{4}{*}{25.46}&\multirow{4}{*}{$\begin{array}{ccc}(3&\bar{13}&10)\\(23&\bar{7}&\bar{16})\\(1&1&1)*\end{array}$}&\multirow{4}{*}{$\begin{pmatrix}3&23&1\\-13&-7&1\\10&-16&1\end{pmatrix}$} \\
\hhline{~---~~~~~~~~}		
&6.03&$\begin{array}{ccc}(19&\bar{1}&0)\\(1&19&0)\\(0&0&1)*\end{array}$&$\begin{pmatrix}19&1&0\\-1&19&0\\0&0&1\end{pmatrix}$ &	&&&	&&& \\
\hline		
\multirow{4}{*}{185}&72.05&$\begin{array}{ccc}(11&\bar{8}&0)\\(8&11&0)\\(0&0&1)*\end{array}$&$\begin{pmatrix}11&8&0\\-8&11&0\\0&0&1\end{pmatrix}$ &	\multirow{4}{*}{137}&\multirow{4}{*}{29.70}&\multirow{4}{*}{$\begin{array}{ccc}(\bar{8}&8&3)\\(3&\bar{3}&16)\\(1&1&0)*\end{array}$}&\multirow{4}{*}{$\begin{pmatrix}-8&3&1\\8&-3&1\\3&16&0\end{pmatrix}$} &	\multirow{4}{*}{147 (49)}&\multirow{4}{*}{16.43}&\multirow{4}{*}{$\begin{array}{ccc}(2&\bar{13}&11)\\(8&\bar{3}&\bar{5})\\(1&1&1)*\end{array}$}&\multirow{4}{*}{$\begin{pmatrix}2&8&1\\-13&-3&1\\11&-5&1\end{pmatrix}$} \\
\hhline{~---~~~~~~~~}		
&17.95&$\begin{array}{ccc}(19&\bar{3}&0)\\(3&19&0)\\(0&0&1)*\end{array}$&$\begin{pmatrix}19&3&0\\-3&19&0\\0&0&1\end{pmatrix}$ &	&&&	&&& \\
\hline		
\multirow{4}{*}{193}&60.51&$\begin{array}{ccc}(12&\bar{7}&0)\\(7&12&0)\\(0&0&1)*\end{array}$&$\begin{pmatrix}12&7&0\\-7&12&0\\0&0&1\end{pmatrix}$ &	\multirow{4}{*}{139}&\multirow{4}{*}{42.18}&\multirow{4}{*}{$\begin{array}{ccc}(\bar{11}&11&6)\\(3&\bar{3}&11)\\(1&1&0)*\end{array}$}&\multirow{4}{*}{$\begin{pmatrix}-11&3&1\\11&-3&1\\6&11&0\end{pmatrix}$} &	\multirow{4}{*}{151}&\multirow{4}{*}{41.27}&\multirow{4}{*}{$\begin{array}{ccc}(5&\bar{14}&9)\\(23&\bar{4}&\bar{19})\\(1&1&1)*\end{array}$}&\multirow{4}{*}{$\begin{pmatrix}5&23&1\\-14&-4&1\\9&-19&1\end{pmatrix}$} \\
\hhline{~---~~~~~~~~}		
&29.49&$\begin{array}{ccc}(19&\bar{5}&0)\\(5&19&0)\\(0&0&1)*\end{array}$&$\begin{pmatrix}19&5&0\\-5&19&0\\0&0&1\end{pmatrix}$ &	&&&	&&& \\
\hline		
\multirow{4}{*}{197}&8.17&$\begin{array}{ccc}(14&1&0)\\(\bar{1}&14&0)\\(0&0&1)*\end{array}$&$\begin{pmatrix}14&-1&0\\1&14&0\\0&0&1\end{pmatrix}$ &	\multirow{4}{*}{153}&\multirow{4}{*}{47.69}&\multirow{4}{*}{$\begin{array}{ccc}(\bar{5}&5&\bar{16})\\(\bar{8}&8&5)\\(1&1&0)*\end{array}$}&\multirow{4}{*}{$\begin{pmatrix}-5&-8&1\\5&8&1\\-16&5&0\end{pmatrix}$} &	\multirow{4}{*}{157}&\multirow{4}{*}{7.93}&\multirow{4}{*}{$\begin{array}{ccc}(1&\bar{13}&12)\\(25&\bar{11}&\bar{14})\\(1&1&1)*\end{array}$}&\multirow{4}{*}{$\begin{pmatrix}1&25&1\\-13&-11&1\\12&-14&1\end{pmatrix}$} \\
\hhline{~---~~~~~~~~}		
&81.83&$\begin{array}{ccc}(15&13&0)\\(\bar{13}&15&0)\\(0&0&1)*\end{array}$&$\begin{pmatrix}15&-13&0\\13&15&0\\0&0&1\end{pmatrix}$ &	&&&	&&& \\
\hline		
\multirow{4}{*}{205}&24.19&$\begin{array}{ccc}(\bar{3}&14&0)\\(\bar{14}&\bar{3}&0)\\(0&0&1)*\end{array}$&$\begin{pmatrix}-3&-14&0\\14&-3&0\\0&0&1\end{pmatrix}$ &	\multirow{4}{*}{163}&\multirow{4}{*}{8.98}&\multirow{4}{*}{$\begin{array}{ccc}(\bar{9}&9&1)\\(1&\bar{1}&18)\\(1&1&0)*\end{array}$}&\multirow{4}{*}{$\begin{pmatrix}-9&1&1\\9&-1&1\\1&18&0\end{pmatrix}$} &	\multirow{4}{*}{163}&\multirow{4}{*}{23.48}&\multirow{4}{*}{$\begin{array}{ccc}(3&\bar{14}&11)\\(25&\bar{8}&\bar{17})\\(1&1&1)*\end{array}$}&\multirow{4}{*}{$\begin{pmatrix}3&25&1\\-14&-8&1\\11&-17&1\end{pmatrix}$} \\
\hhline{~---~~~~~~~~}		
&65.81&$\begin{array}{ccc}(\bar{7}&19&0)\\(\bar{19}&\bar{7}&0)\\(0&0&1)*\end{array}$&$\begin{pmatrix}-7&-19&0\\19&-7&0\\0&0&1\end{pmatrix}$ &	&&&	&&& \\
\hline		
\multirow{4}{*}{221}&84.55&$\begin{array}{ccc}(11&\bar{10}&0)\\(10&11&0)\\(0&0&1)*\end{array}$&$\begin{pmatrix}11&10&0\\-10&11&0\\0&0&1\end{pmatrix}$ &	\multirow{4}{*}{171}&\multirow{4}{*}{12.42}&\multirow{4}{*}{$\begin{array}{ccc}(\bar{4}45)\\(5\bar{5}8)\\(110)*\end{array}$}&\multirow{4}{*}{$\begin{pmatrix}-4&5&1\\4&-5&1\\5&8&0\end{pmatrix}$} &	\multirow{4}{*}{169}&\multirow{4}{*}{55.59}&\multirow{4}{*}{$\begin{array}{ccc}(7&\bar{15}&8)\\(23&\bar{1}&\bar{22})\\(1&1&1)*\end{array}$}&\multirow{4}{*}{$\begin{pmatrix}7&23&1\\-15&-1&1\\8&-22&1\end{pmatrix}$} \\
\hhline{~---~~~~~~~~}		
&5.45&$\begin{array}{ccc}(21&\bar{1}&0)\\(1&21&0)\\(0&0&1)*\end{array}$&$\begin{pmatrix}21&1&0\\-1&21&0\\0&0&1\end{pmatrix}$ &	&&&	&&& \\
\hline		
\multirow{4}{*}{229}&15.19&$\begin{array}{ccc}(\bar{2}&15&0)\\(\bar{15}&\bar{2}&0)\\(0&0&1)*\end{array}$&$\begin{pmatrix}-2&-15&0\\15&-2&0\\0&0&1\end{pmatrix}$ &	\multirow{4}{*}{177}&\multirow{4}{*}{24.55}&\multirow{4}{*}{$\begin{array}{ccc}(\bar{2}&2&\bar{13})\\(\bar{13}&13&4)\\(1&1&0)*\end{array}$}&\multirow{4}{*}{$\begin{pmatrix}-2&-13&1\\2&13&1\\-13&4&0\end{pmatrix}$} &	\multirow{4}{*}{181}&\multirow{4}{*}{29.84}&\multirow{4}{*}{$\begin{array}{ccc}(4&\bar{15}&11)\\(26&\bar{7}&\bar{19})\\(1&1&1)*\end{array}$}&\multirow{4}{*}{$\begin{pmatrix}4&26&1\\-15&-7&1\\11&-19&1\end{pmatrix}$} \\
\hhline{~---~~~~~~~~}		
&74.81&$\begin{array}{ccc}(\bar{13}&17&0)\\(\bar{17}&\bar{13}&0)\\(0&0&1)*\end{array}$&$\begin{pmatrix}-13&-17&0\\17&-13&0\\0&0&1\end{pmatrix}$ &	&&&	&&& \\
\hline		
\multirow{4}{*}{233}&63.22&$\begin{array}{ccc}(\bar{13}&8&0)\\(\bar{8}&\bar{13}&0)\\(0&0&1)*\end{array}$&$\begin{pmatrix}-13&-8&0\\8&-13&0\\0&0&1\end{pmatrix}$ &	\multirow{4}{*}{179}&\multirow{4}{*}{84.55}&\multirow{4}{*}{$\begin{array}{ccc}(\bar{7}&7&9)\\(9&\bar{9}&14)\\(1&1&0)*\end{array}$}&\multirow{4}{*}{$\begin{pmatrix}-7&9&1\\7&-9&1\\9&14&0\end{pmatrix}$} &	\multirow{4}{*}{183 (61)}&\multirow{4}{*}{7.34}&\multirow{4}{*}{$\begin{array}{ccc}(1&\bar{14}&13)\\(9&\bar{4}&\bar{5})\\(1&1&1)*\end{array}$}&\multirow{4}{*}{$\begin{pmatrix}1&9&1\\-14&-4&1\\13&-5&1\end{pmatrix}$} \\
\hhline{~---~~~~~~~~}		
&26.78&$\begin{array}{ccc}(\bar{21}&5&0)\\(\bar{5}&\bar{21}&0)\\(0&0&1)*\end{array}$&$\begin{pmatrix}-21&-5&0\\5&-21&0\\0&0&1\end{pmatrix}$ &	&&&	&&& \\
\hline		
\hline		
\end{tabular}		
\end{table*}		
		
\begin{table*}		
\begin{tabular}{c|c|c|c|c|c|c|c|c|c|c|c}		
\hline		
\hline		
\multicolumn{4}{c|}{[001]}& \multicolumn{4}{c|}{[110]}& \multicolumn{4}{c}{[111]}\\		
\hline		
\multirow{2}{*}{$\Sigma$} &\multirow{2}{*}{$\theta$} &GB &\multirow{2}{*}{CSL}&\multirow{2}{*}{$\Sigma$} &\multirow{2}{*}{$\theta$} &GB &\multirow{2}{*}{CSL}&\multirow{2}{*}{$\Sigma$} &\multirow{2}{*}{$\theta$} &GB &\multirow{2}{*}{CSL}\\		
& &Plane & & & &Plane & & & &Plane & \\		
\hline		
\multirow{4}{*}{241}&29.86&$\begin{array}{ccc}(\bar{4}&15&0)\\(\bar{15}&\bar{4}&0)\\(0&0&1)*\end{array}$&$\begin{pmatrix}-4&-15&0\\15&-4&0\\0&0&1\end{pmatrix}$ &	\multirow{4}{*}{187}&\multirow{4}{*}{36.15}&\multirow{4}{*}{$\begin{array}{ccc}(\bar{13}&13&6)\\(3&\bar{3}&13)\\(1&1&0)*\end{array}$}&\multirow{4}{*}{$\begin{pmatrix}-13&3&1\\13&-3&1\\6&13&0\end{pmatrix}$} &	\multirow{4}{*}{193}&\multirow{4}{*}{51.74}&\multirow{4}{*}{$\begin{array}{ccc}(7&\bar{16}&9)\\(25&\bar{2}&\bar{23})\\(1&1&1)*\end{array}$}&\multirow{4}{*}{$\begin{pmatrix}7&25&1\\-16&-2&1\\9&-23&1\end{pmatrix}$} \\
\hhline{~---~~~~~~~~}		
&60.14&$\begin{array}{ccc}(\bar{11}&19&0)\\(\bar{19}&\bar{11}&0)\\(0&0&1)*\end{array}$&$\begin{pmatrix}-11&-19&0\\19&-11&0\\0&0&1\end{pmatrix}$ &	&&&	&&& \\
\hline		
\multirow{4}{*}{257}&7.15&$\begin{array}{ccc}(16&1&0)\\(\bar{1}&16&0)\\(0&0&1)*\end{array}$&$\begin{pmatrix}16&-1&0\\1&16&0\\0&0&1\end{pmatrix}$ &	\multirow{4}{*}{193}&\multirow{4}{*}{75.29}&\multirow{4}{*}{$\begin{array}{ccc}(6&\bar{6}&11)\\(11&\bar{11}&\bar{12})\\(1&1&0)*\end{array}$}&\multirow{4}{*}{$\begin{pmatrix}6&11&1\\-6&-11&1\\11&-12&0\end{pmatrix}$} &	\multirow{4}{*}{199}&\multirow{4}{*}{14.11}&\multirow{4}{*}{$\begin{array}{ccc}(2&\bar{15}&13)\\(28&\bar{11}&\bar{17})\\(1&1&1)*\end{array}$}&\multirow{4}{*}{$\begin{pmatrix}2&28&1\\-15&-11&1\\13&-17&1\end{pmatrix}$} \\
\hhline{~---~~~~~~~~}		
&82.85&$\begin{array}{ccc}(17&15&0)\\(\bar{15}&17&0)\\(0&0&1)*\end{array}$&$\begin{pmatrix}17&-15&0\\15&17&0\\0&0&1\end{pmatrix}$ &	&&&	&&& \\
\hline		
\multirow{4}{*}{265}&85.02&$\begin{array}{ccc}(12&\bar{11}&0)\\(11&12&0)\\(0&0&1)*\end{array}$&$\begin{pmatrix}12&11&0\\-11&12&0\\0&0&1\end{pmatrix}$ &	\multirow{4}{*}{201}&\multirow{4}{*}{8.09}&\multirow{4}{*}{$\begin{array}{ccc}(\bar{10}&10&1)\\(1&\bar{1}&20)\\(1&1&0)*\end{array}$}&\multirow{4}{*}{$\begin{pmatrix}-10&1&1\\10&-1&1\\1&20&0\end{pmatrix}$} &	\multirow{4}{*}{201 (67)}&\multirow{4}{*}{35.57}&\multirow{4}{*}{$\begin{array}{ccc}(5&\bar{16}&11)\\(9&\bar{2}&\bar{7})\\(1&1&1)*\end{array}$}&\multirow{4}{*}{$\begin{pmatrix}5&9&1\\-16&-2&1\\11&-7&1\end{pmatrix}$} \\
\hhline{~---~~~~~~~~}		
&4.98&$\begin{array}{ccc}(23&\bar{1}&0)\\(1&23&0)\\(0&0&1)*\end{array}$&$\begin{pmatrix}23&1&0\\-1&23&0\\0&0&1\end{pmatrix}$ &	&&&	&&& \\
\hline		
\multirow{4}{*}{269}&75.14&$\begin{array}{ccc}(\bar{13}&10&0)\\(\bar{10}&\bar{13}&0)\\(0&0&1)*\end{array}$&$\begin{pmatrix}-13&-10&0\\10&-13&0\\0&0&1\end{pmatrix}$ &	\multirow{4}{*}{209}&\multirow{4}{*}{23.95}&\multirow{4}{*}{$\begin{array}{ccc}(10&\bar{10}&\bar{3})\\(\bar{3}&3&\bar{20})\\(1&1&0)*\end{array}$}&\multirow{4}{*}{$\begin{pmatrix}10&-3&1\\-10&3&1\\-3&-20&0\end{pmatrix}$} &	\multirow{4}{*}{211}&\multirow{4}{*}{6.84}&\multirow{4}{*}{$\begin{array}{ccc}(1&\bar{15}&14)\\(29&\bar{13}&\bar{16})\\(1&1&1)*\end{array}$}&\multirow{4}{*}{$\begin{pmatrix}1&29&1\\-15&-13&1\\14&-16&1\end{pmatrix}$} \\
\hhline{~---~~~~~~~~}		
&14.86&$\begin{array}{ccc}(\bar{23}&3&0)\\(\bar{3}&\bar{23}&0)\\(0&0&1)*\end{array}$&$\begin{pmatrix}-23&-3&0\\3&-23&0\\0&0&1\end{pmatrix}$ &	&&&	&&& \\
\hline		
\multirow{4}{*}{277}&65.47&$\begin{array}{ccc}(14&\bar{9}&0)\\(9&14&0)\\(0&0&1)*\end{array}$&$\begin{pmatrix}14&9&0\\-9&14&0\\0&0&1\end{pmatrix}$ &	\multirow{4}{*}{211}&\multirow{4}{*}{57.62}&\multirow{4}{*}{$\begin{array}{ccc}(\bar{7}&7&\bar{18})\\(\bar{9}&9&7)\\(1&1&0)*\end{array}$}&\multirow{4}{*}{$\begin{pmatrix}-7&-9&1\\7&9&1\\-18&7&0\end{pmatrix}$} &	\multirow{4}{*}{217}&\multirow{4}{*}{20.32}&\multirow{4}{*}{$\begin{array}{ccc}(3&\bar{16}&13)\\(29&\bar{10}&\bar{19})\\(1&1&1)*\end{array}$}&\multirow{4}{*}{$\begin{pmatrix}3&29&1\\-16&-10&1\\13&-19&1\end{pmatrix}$} \\
\hhline{~---~~~~~~~~}		
&24.53&$\begin{array}{ccc}(23&\bar{5}&0)\\(5&23&0)\\(0&0&1)*\end{array}$&$\begin{pmatrix}23&5&0\\-5&23&0\\0&0&1\end{pmatrix}$ &	&&&	&&& \\
\hline		
\multirow{4}{*}{281}&34.71&$\begin{array}{ccc}(\bar{5}&16&0)\\(\bar{16}&\bar{5}&0)\\(0&0&1)*\end{array}$&$\begin{pmatrix}-5&-16&0\\16&-5&0\\0&0&1\end{pmatrix}$ &	\multirow{4}{*}{219}&\multirow{4}{*}{57.09}&\multirow{4}{*}{$\begin{array}{ccc}(\bar{13}&13&10)\\(5&\bar{5}&13)\\(1&1&0)*\end{array}$}&\multirow{4}{*}{$\begin{pmatrix}-13&5&1\\13&-5&1\\10&13&0\end{pmatrix}$} &	\multirow{4}{*}{219 (73)}&\multirow{4}{*}{48.36}&\multirow{4}{*}{$\begin{array}{ccc}(7&\bar{17}&10)\\(9&\bar{1}&\bar{8})\\(1&1&1)*\end{array}$}&\multirow{4}{*}{$\begin{pmatrix}7&9&1\\-17&-1&1\\10&-8&1\end{pmatrix}$} \\
\hhline{~---~~~~~~~~}		
&55.29&$\begin{array}{ccc}(\bar{11}&21&0)\\(\bar{21}&\bar{11}&0)\\(0&0&1)*\end{array}$&$\begin{pmatrix}-11&-21&0\\21&-11&0\\0&0&1\end{pmatrix}$ &	&&&	&&& \\
\hline		
\multirow{4}{*}{289}&56.14&$\begin{array}{ccc}(\bar{15}&8&0)\\(\bar{8}&\bar{15}&0)\\(0&0&1)*\end{array}$&$\begin{pmatrix}-15&-8&0\\8&-15&0\\0&0&1\end{pmatrix}$ &	\multirow{4}{*}{227}&\multirow{4}{*}{10.77}&\multirow{4}{*}{$\begin{array}{ccc}(\bar{15}&15&2)\\(1&\bar{1}&15)\\(1&1&0)*\end{array}$}&\multirow{4}{*}{$\begin{pmatrix}-15&1&1\\15&-1&1\\2&15&0\end{pmatrix}$} &	\multirow{4}{*}{223}&\multirow{4}{*}{40.73}&\multirow{4}{*}{$\begin{array}{ccc}(6&\bar{17}&11)\\(28&\bar{5}&\bar{23})\\(1&1&1)*\end{array}$}&\multirow{4}{*}{$\begin{pmatrix}6&28&1\\-17&-5&1\\11&-23&1\end{pmatrix}$} \\
\hhline{~---~~~~~~~~}		
&33.86&$\begin{array}{ccc}(\bar{23}&7&0)\\(\bar{7}&\bar{23}&0)\\(0&0&1)*\end{array}$&$\begin{pmatrix}-23&-7&0\\7&-23&0\\0&0&1\end{pmatrix}$ &	&&&	&&& \\
\hline		
\multirow{4}{*}{293}&13.42&$\begin{array}{ccc}(\bar{2}&17&0)\\(\bar{17}&\bar{2}&0)\\(0&0&1)*\end{array}$&$\begin{pmatrix}-2&-17&0\\17&-2&0\\0&0&1\end{pmatrix}$ &	\multirow{4}{*}{233}&\multirow{4}{*}{21.36}&\multirow{4}{*}{$\begin{array}{ccc}(\bar{2}&2&\bar{15})\\(\bar{15}&15&4)\\(1&1&0)*\end{array}$}&\multirow{4}{*}{$\begin{pmatrix}-2&-15&1\\2&15&1\\-15&4&0\end{pmatrix}$} &	\multirow{4}{*}{229}&\multirow{4}{*}{33.25}&\multirow{4}{*}{$\begin{array}{ccc}(5&\bar{17}&12)\\(29&\bar{7}&\bar{22})\\(1&1&1)*\end{array}$}&\multirow{4}{*}{$\begin{pmatrix}5&29&1\\-17&-7&1\\12&-22&1\end{pmatrix}$} \\
\hhline{~---~~~~~~~~}		
&76.58&$\begin{array}{ccc}(\bar{15}&19&0)\\(\bar{19}&\bar{15}&0)\\(0&0&1)*\end{array}$&$\begin{pmatrix}-15&-19&0\\19&-15&0\\0&0&1\end{pmatrix}$ &	&&&	&&& \\
\hline		
\hline		
\end{tabular}		
\end{table*}

\newpage
%

\end{document}